\documentclass[article]{IEEEtran}
\usepackage{amsmath, amssymb}
\newcommand{\Reals}{\mathbb R}
\usepackage{psfrag}
\usepackage{graphicx}
\usepackage{color}
\usepackage{framed}
\begin{document}
\title{Coding Schemes and Asymptotic Capacity of the Gaussian Broadcast and Interference Channels with Feedback} 

\author{\authorblockN{Michael Gastpar, Amos Lapidoth, Yossef Steinberg, and Mich\`ele~Wigger}\thanks{The paper was in part presented at the \emph{2008 ITA Workshop}, in San Diego, USA, at  \emph{ISIT 2008} in Toronto, Canada, and at \emph{ISWCS'11} in Aachen.  The work of M. Gastpar was supported in part by the European ERC
Starting Grant 259530-ComCom.
The work of Y. Steinberg was supported by The Israel Science Foundation,
grant no. 684/11. The work of M.~Wigger was supported in part by the City of Paris, under the program "Emergences". 

M.~Gastpar is with the School of Computer and Communication Sciences at EPFL, Switzerland and with the EECS Department at University of California Berkeley \{email: michael.gastpar@epfl.ch\}.
A.~Lapidoth is with the Department of Information Technology and Electrical Engineering at ETH Zurich, Switzerland \{email: lapidoth@isi.ee.ethz.ch\}. 
Y.~Steinberg is with the Department of Electrical Engineering at the Technion---Israel Institute of Technology \{email: ysteinbe@ee.technion.ac.il\}.
M.~Wigger is with the Communications and Electronics Department, at
Telecom ParisTech, France \{email: michele.wigger@telecom-paristech.fr\}.}}

\maketitle

\newtheorem{remark}{Remark}
\newtheorem{note}{Note}
\newtheorem{defn}{Definition}
\newtheorem{example}{Example}
\newtheorem{ex1c}{Example 1, continued}

\newtheorem{theorem}{Theorem}
\newtheorem{lemma}{Lemma}
\newtheorem{proposition}[theorem]{Proposition}
\newtheorem{corollary}[theorem]{Corollary}
\newtheorem{implication}{Implication}
\newtheorem{algorithm}{Algorithm}

\newcommand{\BC}{\textnormal{BC}}
\newcommand{\BCFB}{\textnormal{BC}}
\newcommand{\BCNoisy}{\textnormal{BC,Noisy}}
\newcommand{\IC}{\textnormal{IC}}
\newcommand{\snr}{\textnormal{SNR}}
\newcommand{\capa}{\mathcal{C}}
\newcommand{\bfpi}{\boldsymbol{\pi}}

\newcommand{\inner}[2]{\left\langle{#1},{#2}\right\rangle}
\newcommand{\trace}[1]{\operatorname{tr}\left(#1\right)} 

\newcommand{\Normal}[2]{\mathcal{N}\!\left({#1},{#2}\right)}
\newcommand{\mat}[1]{\mathsf{#1}}
\newcommand{\vect}[1]{\mathbf{#1}} 
\newcommand{\E}[2][]{\textnormal{\textsf{E}}_{#1}\!\left[#2\right]} %
\newcommand{\Var}[1]{\textnormal{\textsf{Var}}\!\left({#1}\right)} 
\newcommand{\Cov}[2]{\textnormal{\textsf{Cov}}\!\left[{#1},{#2}\right]} 
\newcommand{\trans}[1]{#1^{\textnormal{\textsf{\tiny T}}}} 
\newcommand{\bfY}{\vect{Y}}
\newcommand{\bfX}{\vect{X}}
\newcommand{\bfZ}{\vect{Z}}
\newcommand{\set}[1]{\mathcal{#1}}   

\newcommand{\zminus}{\zeta_{-1}}
\newcommand{\zplus}{\zeta_{+1}}

\newcommand{\CHiSNR}{C_{\textnormal{Hi-SNR}}}
\newcommand{\mw}[1]{{\color{blue}{#1}}}

\begin{abstract}
A coding scheme is proposed for the memoryless Gaussian broadcast channel with correlated noises and feedback. For all noise correlations other than $\pm 1$, the gap between the sum-rate the scheme achieves and the full-cooperation bound vanishes as the signal-to-noise ratio tends to infinity. When the correlation coefficient is $-1$, the gains afforded by feedback are unbounded  and the prelog is doubled. When the correlation coefficient is $+1$ we demonstrate a dichotomy: If the noise variances are equal, then feedback is useless, and otherwise, feedback affords unbounded rate gains and doubles the prelog. The unbounded feedback gains, however, require perfect (noiseless) feedback. When the feedback links are noisy the feedback gains are bounded, unless the feedback noise decays to zero sufficiently fast with the signal-to-noise ratio. 

Extensions to more receivers are also discussed as is the memoryless Gaussian interference channel with feedback.
  \end{abstract}

\section{Introduction}

Among Shannon's most elegant results is that feedback cannot increase
the capacity of a memoryless point-to-point channel.  Feedback can,
however, increase the capacity region of various memoryless
multi-terminal networks such as the 
multiple-access channel (MAC) and the broadcast channel (BC).  Exact
expressions for the feedback capacities are known only for special
networks, e.g., the memoryless Gaussian MAC 
\cite{Ozarow:79}, \cite{Ozarow:84}.

This paper considers the memoryless Gaussian BC with feedback.  In the standard
setting, the signals at the different receivers are corrupted by
independent noises.  For this setting, Ozarow and
Leung~\cite{Ozarow:79,OzarowL:84} showed that, indeed, feedback can
enlarge the capacity region, though the exact capacity region with
feedback remains to date unknown.  A natural benchmark is the
``full-cooperation bound,'' where the receivers are allowed to
cooperate. This turns the BC into a (single-input
multiple-output) point-to-point channel whose capacity is well known
and is not increased by feedback.  Not surprisingly, even with
feedback, the full-cooperation bound is generally not attainable.

In this paper we consider the case where the noises at the receivers
are correlated, e.g., due to a common external interference.  In
the absence of feedback, such correlation does
not impact the capacity region, because the latter depends only on the
marginal channels. In the presence of feedback, however, the
correlation is key.

Positively correlated noises were already considered by Ozarow and
Leung. Willems and van der Meulen~\cite{WillemsM:81} extended Ozarow
and Leung's scheme to negatively correlated noises. They also observed
that when the two noises are of equal variances, the sum-rate achieved
by the Ozarow-Leung scheme decreases as the correlation increases. (In
the limiting case of fully correlated noises of equal variances,
feedback does not increase capacity at all.)  Willems and van der
Meulen's observation inspired the current investigation: We wanted to
see whether this observation is an artifact of the specific scheme
they studied or whether it applies to the capacity region. And we
wanted to see how crucial is the assumption that the noises are of
equal variance (very much so!).

In this paper, we present a novel coding scheme and show that---in the
high signal-to-noise ratio (SNR) limit---it achieves the
full-cooperation bound for all noise correlations~$\rho_z$ satisfying
$-1<\rho_z<1$. Consequently, the sum-rate capacity with noise-free
feedback, $C_{\BC, \Sigma}$, satisfies
\begin{framed}\begin{IEEEeqnarray}{rCl}
 \lefteqn{ \lim_{P \to \infty}\Bigg[
  C_{\BC, \Sigma} -  \frac{1}{2} \log \left(
     1+\frac{P(\sigma_1^2+\sigma_2^2-2\rho_z
       \sigma_1\sigma_2)}{\sigma_1^2\sigma_2^2(1-\rho_z^2)}\right)
   \Bigg] } \quad\nonumber\\
  &= &0, \hspace{7cm}\\ 
  && \hspace{3cm}\quad \rho_z \in (-1,1),\hspace{1cm}\nonumber
\end{IEEEeqnarray}
\end{framed}
where $P$ denotes the transmit power and $\sigma_1^2, \sigma_2^2>0$ the noise variances at the two receivers.

The case where $|\rho_z|=1$ is special.  As already mentioned, when
$\rho_z=1$ and the noises are of equal variance, feedback has no
effect on capacity because in this case the BC is merely a
point-to-point channel in disguise.  But when $\rho_z=1$ and the two
noise variances differ, or when $\rho_z=-1$ the full-cooperation bound
is infinite for all $\text{SNR} > 0$, and it is thus useless.  An
alternative upper bound is the sum of the single-user capacities of
the marginal channels to each receiver.

\emph{Prima Facie}, it seems that this upper bound is
completely out of reach because it ignores the tension between
the users. Nevertheless, perhaps surprisingly, we show that
in the high-SNR limit, this upper bound becomes achievable.  More
precisely, for noise correlation $\rho_z=-1,$ as well as for noise
correlation $\rho_z=1$ provided that $\sigma_1^2\not=\sigma_2^2$, 
\begin{framed}
\begin{IEEEeqnarray}{rCl}
 \lefteqn{ \lim_{P \to \infty}\Bigg[
  C_{\BC, \Sigma} -  \left( \frac{1}{2}\log\left(1+ \frac{P}{\sigma_1^2}\right) +\frac{1}{2}
    \log\left(1+ \frac{P}{\sigma_2^2}\right) \right)\Bigg] } \; \nonumber\\
  &= &0, \hspace{7.5cm}\\&&\Bigl(\text{when $\rho_z = -1$, or when $\rho_{z} = 1$ and
    $\sigma_1^2\not=\sigma_2^2$} \Bigr).\nonumber
\end{IEEEeqnarray}
\end{framed}
Without feedback, the sum-rate capacity is $\frac{1}{2}\log_2( 1 +
P/(\min\{ \sigma_1^2, \sigma_2^2\}))$, so, for such noise correlations, feedback
asymptotically doubles the sum-rate capacity in the high-SNR regime
and the \emph{prelog} becomes
\begin{framed}
\begin{IEEEeqnarray}{rCl}
 \lefteqn{ 
\varlimsup_{P\to \infty} \frac{C_{\textnormal{BC}, \Sigma}}{\frac{1}{2}\log P} = 2}\\
&&\Bigl(\text{when $\rho_z = -1$, or when $\rho_{z} = 1$ and
    $\sigma_1^2\not=\sigma_2^2$} \Bigr).\nonumber
\end{IEEEeqnarray}
\end{framed}

To put this result in context, it is important to note that, although
feedback does provide capacity gains in many networks, these gains are
typically modest and bounded in the SNR.
By contrast, the present paper exhibits instances of networks
where the capacity gains afforded by feedback are unbounded in the
SNR.  To the best of our knowledge, these are the first examples of such
large feedback gains. Such examples were first reported in
\cite{GastparW:10-1} for the two-user memoryless BC and for the symmetric
two-user memoryless interference channel (IC) where the individual noise
sequences corrupting the outputs at the two receivers are {perfectly
  anti-correlated}.  In the meantime, other 
networks have been found where feedback affords unbounded
capacity gains; see \cite{SuhTse:09, SuhTse:10} (based on
the scheme proposed in \cite{Tuninetti:07}) for the two-user Gaussian IC
when the noise sequences are {independent}.  Multiplicative gains for
the Gaussian IC with independent noises at moderate SNR were already
reported in \cite[Section~VI-B]{Kramer:02}.

There are several important ensuing questions concerning the special
case of fully correlated noises. For example, we show that even if the
correlation is not perfect but tends to one (or minus one) at least
inversely proportionally to the SNR, we also obtain the same
asymptotic capacity gain.  Another question concerns the case where
the feedback is noisy.  We show that when the feedback links are
corrupted by independent  Gaussian noise sequences, then---irrespective of the
positive feedback-noise variances and of the correlation of the
forward noise-sequences---the prelog of the two-user Gaussian BC setup
equals one (as in the absence of feedback). The proof of this result
is based on a genie argument inspired by the work of Kim, Lapidoth,
and Weissman \cite{KimLapidothWeissman:06}.  

Finally, we consider the $K$-receivers memoryless Gaussian BC with $K > 2$, where no
two of the Gaussian noise sequences corrupting the $K$ received signals
are of equal variance; none of the noise sequences is of zero variance;
and the covariance matrix of the $K$ noises is of rank $1$.  For this
setup, our proposed coding scheme proves the achievability of a prelog
of $K$. For a related recent result see \cite{ArdestMF:11}.

The second network we consider is the two-user scalar memoryless Gaussian IC with
noise-free \emph{one-sided} feedback where each of the two
transmitters communicates with a different intended receiver, and each
transmitter observes feedback from its corresponding receiver
only. Our proposed coding scheme proves that if the noise sequences at
the two receivers are perfectly anticorrelated or perfectly
correlated, then, for most channel gains, noise-free feedback doubles
the prelog from 1 to 2.  Noise-free feedback thus approximately
doubles the sum-rate capacity at high SNR and thus provides unbounded
capacity gains.  (When the interference channel is symmetric, the
prelog~$2$ result can also be shown using a slight generalization (to
account for the correlation between the noise sequences) of Kramer's
memoryless LMMSE-scheme \cite{Kramer:02}.)

Previously, a prelog of 2 was known to be achievable for the two-user
scalar Gaussian IC only when the two transmitters (or the two receivers)
could \emph{fully} cooperate \cite{DevroyeSharif:07} in the sense that
both transmitters could compute their channel inputs as a function of
both messages. Our result shows that \emph{limited} cooperation
through feedback can be sufficient.

For the two-user Gaussian IC we do not consider noisy feedback. Rate-limited feedback for this setup has recently been studied in \cite{VahidSA:11}.

We conclude this section with some notation and a brief outline of the
rest of the paper.  Throughout the paper logarithms are base 2, and for
convenience we define $-\log 0 =\infty$. We use the shorthand
notation $\log^+(x)$ for $\max\{0, \log(x)\}$. Also, we denote by
$A^n$ and $a^n$ the tuple of random variables $A_1, \ldots, A_n$ and
their realizations $a_1, \ldots, a_n$, respectively.  The set of real
numbers is denoted by $\Reals$, the set of positive real numbers
by $\Reals^+$, and the set of positive integers by $\mathbb{Z}^+$. The
abbreviation \emph{IID} stands for \emph{independent and identically
  distributed}.
 
 The paper is organized as follows. 
In the following Section~\ref{sec:pfb} we study the two-user Gaussian BC with noise-free or noisy feedback; 
in Section~\ref{sec:KBC} we study the $K$-user Gaussian BC with noise-free feedback; and in Section~\ref{sec:inter} the two-user Gaussian IC with noise-free feedback. 

\section{Two-User Broadcast Channel}\label{sec:pfb}
\subsection{Setup}\label{sec:model1}

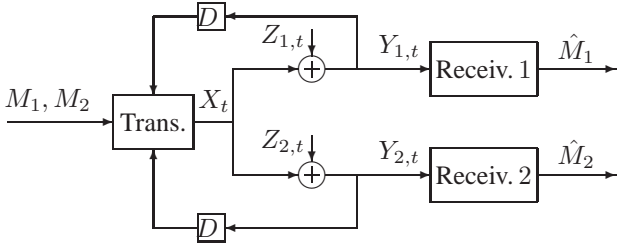
\begin{figure}

  \begin{center}
  \setlength{\unitlength}{1pt}
  \begin{picture}(230,100)(-30,50)

    \put (-30, 95) {\vector (1, 0) {40} }
    \put (-25, 98) {\makebox (20,10) {$M_1, M_2$}}

    \put (10, 85) {\framebox (30,20) {{ Trans. }}}

    \put (40, 95) {\line (1, 0) {15} }
    \put (43, 98) {\makebox (10,10) {$X_t$}}

    \put (55, 75) {\line (0, 1) {40} }

    \put (55, 115) {\vector (1, 0) {25} }
   \put (90, 115) {\vector (1, 0) {40} }
    \put (113, 118) {\makebox (10,10) {$Y_{1,t}$}}
    \put (85, 115) {\circle {10} }
    \put (82, 115) {\line (1, 0) {6} }
    \put (85, 112) {\line (0, 1) {6} }
    \put (85, 130) {\vector (0, -1) {10} }
    \put (69, 122) {\makebox (10,10) {$Z_{1,t}$}}

    \put (102, 115) {\line (0, 1) {20} }
    \put (102, 135) {\vector (-1, 0) {50} }
    \put (42, 130) {\framebox (10,10) {{$D$ }}}
    \put (42, 135) {\line (-1, 0) {17} }
    \put (25, 135) {\vector (0, -1) {30} }

    \put (55, 75) {\vector (1, 0) {25} }
    \put (90, 75) {\vector (1, 0) {40} }
    \put (113, 78) {\makebox (10,10) {$Y_{2,t}$}}
    \put (85, 75) {\circle {10} }
    \put (82, 75) {\line (1, 0) {6} }
    \put (85, 72) {\line (0, 1) {6} }
    \put (85, 90) {\vector (0, -1) {10} }
    \put (69, 82) {\makebox (10,10) {$Z_{2,t}$}}

    \put (102, 75) {\line (0, -1) {20} }
    \put (102, 55) {\vector (-1, 0) {50} }
    \put (42, 50) {\framebox (10,10) {{$D$ }}}
    \put (42, 55) {\line (-1, 0) {17} }
    \put (25, 55) {\vector (0, 1) {30} }

    \put (130, 105) {\framebox (40,20) {{Receiv. }$\!1$}}
    \put (170, 115) {\vector (1, 0) {30} }
    \put (180, 118) {\makebox (10,10) {$\hat{M}_1$}}

    \put (130, 65) {\framebox (40,20) {{Receiv. }$\!2$}}
    \put (170, 75) {\vector (1, 0) {30} }
    \put (180, 78) {\makebox (10,10) {$\hat{M}_2$}}

  \end{picture}
  \end{center}
  \caption{The two-user Gaussian BC with noise-free feedback.}
  \label{Fig-2broadcast}
\end{figure}

We consider the real, scalar, memoryless Gaussian BC. 
Denoting the time-$t$ transmitted symbol
by $x_{t} \in \Reals$ and the time-$t$ received symbols by $Y_{1,t}$
and $Y_{2,t}$,
\begin{subequations}\label{eq:BCY}
\begin{eqnarray}
Y_{1,t} & = & x_{t} +Z_{1,t}, \label{eq:Y1}
\end{eqnarray}
\begin{eqnarray}
Y_{2,t} & = & x_{t} + Z_{2,t},\label{eq:Y2}
\end{eqnarray}
\end{subequations}
where the sequence of noise pairs $\{( Z_{1,t}, Z_{2,t})\}$ is drawn
IID according to a
 centered Gaussian distribution of covariance matrix
\begin{equation}\label{eq:matK}
\mat{K}_z = \begin{pmatrix}\sigma_1^2  & \rho_z \sigma_1\sigma_2\\ \rho_z \sigma_1 \sigma_2 & \sigma_2^2\end{pmatrix}.
\end{equation}
We assume that both noise variances $\sigma_1^2, \sigma_2^2$ are
strictly positive, and we denote their positive roots $\sigma_1, \sigma_2$.

%

The transmitter wishes to send Message~$M_{1}$ to Receiver~1 and an
independent message $M_2$ to Receiver~2.  The messages $M_1$ and $M_2$
are assumed to be uniformly distributed over the sets
$\mathcal{M}_1\triangleq\{1,\ldots, \lfloor 2^{n R_1} \rfloor\}$ and
$\mathcal{M}_2\triangleq \{1,\ldots, \lfloor 2^{n R_2} \rfloor\}$,
where $n$ denotes the blocklength and $R_1$ and $R_2$ the respective
rates of transmission.

\begin{figure}
  \begin{center}
  \setlength{\unitlength}{1pt}
  \begin{picture}(230,100)(-30,50)
    \put (-35, 95) {\vector (1, 0) {35} }
    \put (-33, 98) {\makebox (20,10) {$M_1, M_2$}}
    \put (00, 85) {\framebox (30,20) {{Trans.}}}
    \put (30, 95) {\line (1, 0) {25} }
    \put (36, 98) {\makebox (10,10) {$X_t$}}
    \put (55, 75) {\line (0, 1) {40} }
    \put (55, 115) {\vector (1, 0) {25} }
   \put (90, 115) {\vector (1, 0) {40} }
    \put (113, 118) {\makebox (10,10) {$Y_{1,t}$}}
    \put (85, 115) {\circle {10} }
    \put (82, 115) {\line (1, 0) {6} }
    \put (85, 112) {\line (0, 1) {6} }
    \put (85, 130) {\vector (0, -1) {10} }
    \put (89, 122) {\makebox (10,10) {$Z_{1,t}$}}
    \put (102, 115) {\line (0, 1) {20} }
    \put (102, 135) {\vector (-1, 0) {30} }
    \put (35, 130) {\framebox (10,10) {{$D$ }}}
    \put (61, 135) {\line (-1, 0) {15} }
    \put (67, 135) {\circle {10} }
    \put (64, 135) {\line (1, 0) {6} }
    \put (67, 132) {\line (0, 1) {6} }
    \put (68, 150) {\vector (0, -1) {10} }
    \put (74, 142) {\makebox (10,10) {$W_{1,t}$}}
    \put (25,135) {\line (1,0) {10}}
    \put (25, 135) {\vector (0, -1) {30} }
    \put (55, 75) {\vector (1, 0) {25} }
    \put (90, 75) {\vector (1, 0) {40} }
    \put (113, 78) {\makebox (10,10) {$Y_{2,t}$}}
    \put (85, 75) {\circle {10} }
    \put (82, 75) {\line (1, 0) {6} }
    \put (85, 72) {\line (0, 1) {6} }
    \put (85, 90) {\vector (0, -1) {10} }
    \put (89, 82) {\makebox (10,10) {$Z_{2,t}$}}
	\put (102, 75) {\line (0, -1) {20} }
    \put (102, 55) {\vector (-1, 0) {30} }
    \put (35, 50) {\framebox (10,10) {{$D$ }}}
    \put (61, 55) {\vector (-1, 0) {16} }
        \put (67, 55) {\circle {10} }
    \put (64, 55) {\line (1, 0) {6} }
    \put (67, 52) {\line (0, 1) {6} }
    \put (68, 40) {\vector (0, 1) {10} }
    \put (74, 36) {\makebox (10,10) {$W_{2,t}$}}
    \put (25,55) {\line (1,0) {10}}
    \put (25, 55) {\vector (0, 1) {30} }
    \put (130, 105) {\framebox (40,20) {{Receiv. }$\!1$}}
    \put (170, 115) {\vector (1, 0) {20} }
    \put (180, 118) {\makebox (10,10) {$\hat{M}_1$}}
    \put (130, 65) {\framebox (40,20) {{Receiv. }$\!2$}}
    \put (170, 75) {\vector (1, 0) {20} }
    \put (180, 78) {\makebox (10,10) {$\hat{M}_2$}}
  \end{picture}
  \end{center}
  \caption{The two-user  Gaussian BC with noisy feedback.}
  \label{Fig-2broadcastnoisy}
\end{figure}
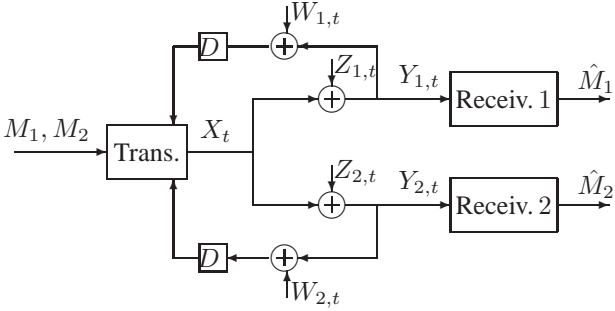

We depict  the scenario with \emph{noise-free feedback} in
Figure~\ref{Fig-2broadcast} and with \emph{noisy feedback} in
Figure~\ref{Fig-2broadcastnoisy}. In the former 
the transmitter learns the outputs $Y_{1,t-1}$ and $Y_{2,t-1}$ after
sending~$X_{t-1}$. It can thus choose its time-$t$ channel input
$X_{t}$ as a
function of both messages and all previous channel outputs:
\begin{equation}\label{eq:power1}
X_{t}=f^{(n)}_{\textnormal{BC},t}\left( M_1, M_2, {Y}_1^{t-1}, {Y}_2^{t-1} \right), \quad t\in\{1,\ldots,n\},
\end{equation}
where the encoding function $f^{(n)}_{\BC,t}$ is of the form
\begin{equation}\label{eq:encdef}
f^{(n)}_{\textnormal{BC},t} \colon \mathcal{M}_1 \times
\mathcal{M}_2\times \mathbb{R}^{t-1} \times \mathbb{R}^{t-1} \to \Reals.
\end{equation}

In the scenario with \emph{noisy feedback} the transmitter, after
sending~$X_{t-1}$, does not learn $Y_{1,t-1}$ and $Y_{2,t-1}$ but
instead learns $V_{1,t-1}$ and $V_{2,t-1}$, which are noisy versions of
$Y_{1,t-1}$ and $Y_{2,t-1}$: 
\begin{IEEEeqnarray*}{rCl}
V_{1,t-1} & = & Y_{1,t-1} +W_{1,t-1}, \\
V_{2,t-1} & = & Y_{2,t-1} +W_{2,t-1},
\end{IEEEeqnarray*}
where the sequence of pairs of feedback noises $\{(W_{1,t},W_{2,t})\}$ is
IID according to a
zero-mean bivariate Gaussian distribution of diagonal\footnote{We do not treat setups with correlated feedback noises
  or setups with feedback noises that are correlated with the forward
  noises.} covariance matrix 
\begin{equation}
\begin{pmatrix}\sigma_{W1}^2 & 0 \\ 0 &
  \sigma_{W2}^2
\end{pmatrix}, \quad \sigma_{W1}, \sigma_{W2} > 0.
\end{equation}
The sequence $\{(W_{1,t},W_{2,t})\}$ is assumed to be independent of
the messages $(M_1,M_2)$ and the noise sequences on the forward path
$\{(Z_{1,t},Z_{2,t})\}$.  In this scenario the transmitter chooses its
time-$t$ channel input $X_{t}$ as
\begin{equation}
X_t = f_{\BC\textnormal{Noisy},t}^{(n)}\left(M_1,M_2, V_{1}^{t-1}, V_{2}^{t-1}\right), \quad t\in
\{1,\ldots, n\},
\end{equation}
where the encoding function is of the form
\begin{equation}
f_{\BC\textnormal{Noisy},t}^{(n)} \colon \mathcal{M}_{1}\times
\mathcal{M}_2 \times \mathbb{R}^{t-1}\times \mathbb{R}^{t-1} \to \Reals.
\end{equation}

In both scenarios, the channel inputs  are subject to an
expected average block-power constraint $P>0$. Thus, 
we only allow encoding functions
$\left\{f^{(n)}_{\textnormal{BC},t}\right\}_{t=1}^{n}$ or
$\left\{f_{\BC\textnormal{Noisy},t}^{(n)}\right\}_{t=1}^n$  for which
\begin{equation}
\frac{1}{n} \E{  \sum_{t=1}^nX_t^2  } \leq P. \label{eq:power}
\end{equation}

Receiver~$k \in \{1,2\}$ decodes its desired message $M_k$ based on its observed channel output sequence $Y_{k}^n$.  That is, it produces the estimate
\begin{eqnarray}
\hat{M}_k&= &\phi_k^{(n)}({Y}_{k}^n),\qquad k\in\{1,2\}, 
\end{eqnarray}
using some decoding function 
\begin{IEEEeqnarray}{rCl} 
\phi_k^{(n)} &\colon& \mathbb{R}^{n} \rightarrow \{1,\ldots, \lfloor 2^{n R_k}
\rfloor\} ,\qquad k\in\{1,2\}.\label{eq:decdef1}
\end{IEEEeqnarray}

For the scenario with noise-free feedback, a rate pair $(R_1,R_2)$ is said to be achievable  if for every
block-length $n$ there exists a set of $n$ encoding functions
$\{f_{\textnormal{BC},t}^{(n)}\}_{t=1}^n$ satisfying the power constraint~\eqref{eq:power}
and two decoding functions $\phi_1^{(n)}$ and $\phi_2^{(n)}$ 
such that
\begin{equation*}
\lim_{n\rightarrow \infty}\text{Pr}\left[(M_1,M_2)\neq
(\hat{M}_1,\hat{M}_2)\right] =0.
\end{equation*}
The closure of the set of all achievable rate pairs $(R_1,R_2)$ is the
\emph{capacity region}. The supremum of the sum $R_1+R_2$ over all
achievable rate pairs $(R_1,R_2)$ is the \emph{sum-rate capacity},
which is denoted
$C_{\BC,\Sigma}(P,\sigma_1^2, \sigma_2^2, \rho_z)$.

For the scenario with noisy feedback, achievable rates, the capacity
region, and the sum-rate capacity are defined analogously but using
the encoding functions
$\{f_{\textnormal{BCNoisy},t}^{(n)}\}_{t=1}^n$. The sum-rate capacity
with noisy feedback is denoted by
$C_{\BC\textnormal{Noisy},\Sigma}(P,\sigma_1^2, \sigma_2^2, \rho_z,
\sigma_{W1}^2, \sigma_{W2}^2)$.

The \emph{prelog}, characterizes the logarithmic growth of the
sum-rate capacity at high SNR. In the scenario with noise-free
feedback it is defined as
\begin{eqnarray}
  \varlimsup_{P \rightarrow \infty} \frac{C_{\BCFB,\Sigma}(P,\sigma_1^2, \sigma_2^2, \rho_z)}{\frac{1}{2} \log ( 1 + P )}
\end{eqnarray}
and in the scenario with noisy feedback as
\begin{eqnarray}
  \varlimsup_{P \rightarrow \infty} \frac{C_{\BC\textnormal{Noisy},\Sigma}(P,\sigma_1^2, \sigma_2^2, \rho_z, \sigma_{W1}^2, \sigma_{W2}^2)}{\frac{1}{2} \log ( 1 + P )}.
\end{eqnarray}

\subsection{Main Results}
\label{sec:main}
Our results depend on whether or not the channel is physically degraded. The
Gaussian BC is physically degraded whenever
\begin{equation}\label{eq:pd}
\rho_z = \frac{\sigma_1}{\sigma_2}\;\;\; \text{or} \;\; \;\rho_{z} = \frac{\sigma_2}{\sigma_1}.
\end{equation}
For example, it is physically degraded
when $\rho_z=1$ and $\sigma_1 =\sigma_2$, in which case the
receivers observe the same sequence.
When the Gaussian BC is physically degraded, feedback does not increase
capacity \cite{ElGamal:78}, and thus \cite{Bergmans:74, Cover:72}
\begin{IEEEeqnarray}{rcl}\label{eq:BCphysdeg}
 C_{\BCFB,\Sigma}(P, \sigma_1^2, \sigma_2^2, \rho_z)&=&C_{\BC\textnormal{Noisy},\Sigma}(P, \sigma_1^2, \sigma_2^2, \rho_z, \sigma_{W1}^2, \sigma_{W2}^2)  \nonumber \\ 
&=& \frac{1}{2}\log\left(1+\frac{P}{\min\{\sigma_1^2, \sigma_2^2\}}\right),
\end{IEEEeqnarray}
irrespective of $\sigma_{W1}^2, \sigma_{W2}^2$ and $\rho_z$. If the
channel is not physically degraded, then $C_{\BCFB, \Sigma}$ and
$C_{\BC\textnormal{Noisy},\Sigma}$ are, in general, unknown, and
bounds are called for.

We first present our results for noise-free feedback. For this
scenario Theorem~\ref{th:highSNR} reveals the high-SNR asymptotic
sum-rate capacity. As we shall see, feedback strictly improves this
asymptote whenever the BC is not physically degraded.

We shall express the asymptotic behavior using the function
$\CHiSNR(P, \sigma_1^2, \sigma_2^2, \rho_z)$ whose
definition depends on whether the BC is physically degraded and on
whether $\rho_z$ is strictly between $-1$ and $+1$:
\begin{defn}\label{def:highC}
Define $\CHiSNR (P, \sigma_1^2, \sigma_2^2, \rho_z)$ as follows.
\begin{itemize}
\item
For channels that are physically degraded, 
\begin{IEEEeqnarray}{rCl}\label{eq:defhi1}
\CHiSNR(P, \sigma_1^2, \sigma_2^2, \rho_z)\triangleq\frac{1}{2} \log \left(1+ \frac{P}{\min\{\sigma_1^2, \sigma_2^2\}} \right).\nonumber \\
\end{IEEEeqnarray}
\item For channels that are not physically degraded and for which
  $\rho_z\in\{-1,1\}$,
\begin{IEEEeqnarray}{rCl}
\CHiSNR(P, \sigma_1^2, \sigma_2^2, \rho_z)\triangleq\frac{1}{2} \log\frac{P}{\sigma_1^2} + \frac{1}{2} \log \frac{ P}{\sigma_2^2}. \IEEEeqnarraynumspace
\end{IEEEeqnarray}
\item For channels that are not physically degraded and where the
  noises are only partially correlated, i.e., $\rho_z \in (-1,1)$ and
  $\rho_z \notin\left\{ \frac{\sigma_1}{\sigma_2},
    \frac{\sigma_2}{\sigma_1}\right\}$,\footnote{Notice that for
    physically degraded channels with partially correlated noises,
    i.e., for $\rho_z \in(-1,1)$ and $\rho_z \in\left\{
      \frac{\sigma_1}{\sigma_2}, \frac{\sigma_2}{\sigma_1}\right\}$,
    the definitions in \eqref{eq:defhi1} and \eqref{eq:defhi3}
    coincide.}
\begin{IEEEeqnarray}{rCl}\label{eq:defhi3}
\lefteqn{
\CHiSNR(P, \sigma_1^2, \sigma_2^2, \rho_z)}\qquad \nonumber \\ & \triangleq & \frac{1}{2}\log\left(\frac{P ( \sigma_1^2+\sigma_2^2- 2\rho_z \sigma_1\sigma_2)}{\sigma_1^2\sigma_2^2(1-\rho_z^2)}\right) .
\end{IEEEeqnarray}
\end{itemize}
\end{defn}
\begin{theorem}
\label{th:highSNR}
For all $\sigma_1, \sigma_2>0$ and $\rho_z\in[-1,1]$
\begin{equation}\label{eq:highSNR}\lim_{P\to \infty}
\big(C_{\BC,\Sigma}(P,\sigma_1^2, \sigma_2^2, \rho_z)- \CHiSNR (P, \sigma_1^2, \sigma_2^2, \rho_z)\big) =0. 
\end{equation}
\end{theorem}
\begin{proof}
When the channel is physically degraded, 
the result follows  from \eqref{eq:BCphysdeg}. 

When $\rho_z\in\{-1,1\}$ and the BC is not physically degraded,
the desired rates are achieved by a novel scheme that we describe in
Section~\ref{sec:proof-awgn-bc} ahead (see Corollary~\ref{lem:K20} in
Section~\ref{sec:highSNR}). The converse for this case follows by
applying the cut-set bound with two cuts, one between the transmitter
and each of the two receivers:
\begin{IEEEeqnarray}{rCl}
R_1+ R_ 2 & \leq&  \max_{X \colon \E{X^2} \leq P} \{I(X;Y_1) +I(X;Y_2)\}\nonumber\\
& = & \frac{1}{2} \log\left( 1+ \frac{P}{\sigma_1^2}\right)+ \frac{1}{2} \log\left( 1+ \frac{P}{\sigma_2^2}\right),
\end{IEEEeqnarray}
where the equality follows because a Gaussian law maximizes the differential entropy under a variance constraint \cite{CoverT:06}.

When  $\rho_z\in(-1,1)$ and the channel is not physically degraded,
the achievability is demonstrated using our scheme of 
Section~\ref{sec:proof-awgn-bc} (see Corollary~\ref{cor:highSNR} in
Section~\ref{sec:highSNR}); the converse follows by applying the
cut-set bound with a single cut separating the transmitter from the two
receivers:
\begin{IEEEeqnarray}{rCl}
R_1+ R_ 2 & \leq&  \max_{X \colon \E{X^2} \leq P} I(X;Y_1,Y_2) \nonumber\\
& \leq & \frac{1}{2} \log\left( 1+ \frac{P(\sigma_1^2+\sigma_2^2- 2\sigma_1\sigma_2 \rho_z )}{ \sigma_1^2\sigma_2^2(1-\rho_z^2) }\right). \IEEEeqnarraynumspace
\end{IEEEeqnarray}

\end{proof}

In general, the previously proposed schemes in~\cite{Ozarow:79},
\cite{OzarowL:84}, and \cite{ArdestMF:11} cannot achieve the high-SNR
asymptotic sum-rate capacity $\CHiSNR$: The scheme in
\cite{Ozarow:79}, \cite{OzarowL:84}, for example, achieves
$\CHiSNR$ when $\rho_z \leq 0$ but not when
$\rho_z>0$. And the scheme in \cite[Theorem~2]{ArdestMF:11} achieves
$\CHiSNR$ when $\rho_z=0$ and $\sigma_1=\sigma_2$,
but it does not apply when $\sigma_1 \neq \sigma_2$.

\begin{note} If $|\rho_z| < 1$ and the feedback links are
  noise-free, then the high-SNR sum-rate capacity 
  $\CHiSNR(P,\sigma_1^2, \sigma_2^2, \rho_z)$
  is as though the two receivers could fully cooperate in their decoding.
  If $\rho_z\in\{-1,1\}$ and the channel is not physically degraded,
  then the high-SNR sum-rate capacity is as though there were a separate
  (non-interfering) link from the transmitter to each of the 
  receivers and the transmitter could communicate with full power
  $P$ over each of these links.
\end{note}

\begin{figure}[tbp]

 \psfrag{5}[r][r]{}
 \psfrag{10}[r][r]{\scriptsize$10$}
 \psfrag{15}[r][r]{}
 \psfrag{20}[r][r]{\scriptsize$20$}
 \psfrag{25}[r][r]{}
  \psfrag{30}[r][r]{\scriptsize$30$}
 \psfrag{40}[r][r]{\scriptsize$40$}
 \psfrag{35}[r][r]{}
 \psfrag{1}[c][c]{}
  \psfrag{0.6}[c][c]{}
 \psfrag{0.4}[c][c]{\scriptsize$0.4$}
 \psfrag{0.2}[c][c]{}
 \psfrag{0}[c][c]{\scriptsize$0$}
 \psfrag{0.8}[c][c]{\scriptsize$0.8$}
  \psfrag{-1}[c][c]{}
  \psfrag{-0.6}[c][c]{}
 \psfrag{-0.4}[c][c]{\scriptsize$-0.4$}
 \psfrag{-0.2}[c][c]{}
 \psfrag{-0.8}[c][c]{\scriptsize$-0.8$}
 \psfrag{rz}[t][t]{$\scriptsize \rho_z$}
 \psfrag{gamma}[r][r]{$\scriptsize \gamma(\rho_z)$}
   \includegraphics[width=\columnwidth]{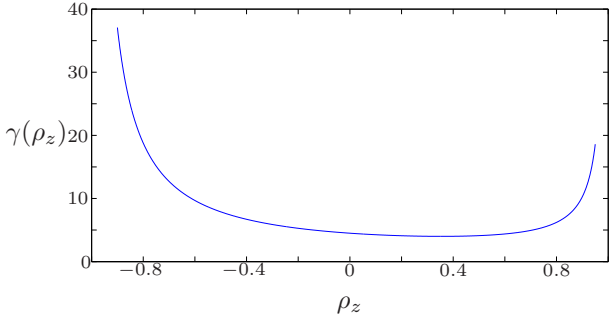}
     \caption{The function $\gamma(\rho_z)$ is plotted over $\rho_z\in[-0.9, 0.95]$  for $\sigma_1^2=2$ and $\sigma_2^2=0.25$. The minimum is at $\sqrt{1/8}\approx 0.3536$, and the function is strictly decreasing over $\big(-1,\sqrt{1/8}\big)$ and strictly increasing over $\big(\sqrt{1/8}, 1\big)$.} 
   \label{fig:2b}
 \end{figure}

\begin{note}
Given $\sigma_2,  \sigma_1>0$, define the \emph{power offset} 
\begin{IEEEeqnarray}{rCl}
\gamma \colon (-1,1)&\to& \Reals^+ \\
 \rho_z &\mapsto& \frac{\sigma_1^2 +\sigma_2^2 -2\rho_z \sigma_1\sigma_2}{\sigma_1^2 \sigma_2^2 (1-\rho_z^2)} 
\end{IEEEeqnarray}
so
\begin{equation*}
  \CHiSNR(P,\sigma_1^2, \sigma_2^2, \rho_z) =
  \frac{1}{2} \log \bigl( P \, \gamma(\rho_z) \bigr), \quad |\rho_z| < 1.
\end{equation*}
Notice that $\gamma(\rho_z)\to \infty$ as $\rho_z\to-1$. Also, if
$\sigma_1\neq \sigma_2$, then $\gamma(\rho_z)\to\infty$ as $\rho_z \to
+1$. Moreover, $\gamma(\rho_z)$ is strictly decreasing for $\rho_z
\in\left(-1, \min\left\{ \frac{\sigma_1}{\sigma_2},
    \frac{\sigma_2}{\sigma_1}\right\}\right)$ and strictly increasing
for $\rho_z\in \left(\min\left\{ \frac{\sigma_1}{\sigma_2},
    \frac{\sigma_2}{\sigma_1}\right\}, 1\right)$.  The power offset
is thus minimal at $\rho_z = \min\left\{
  \frac{\sigma_1}{\sigma_2}, \frac{\sigma_2}{\sigma_1}\right\}$ where
$\gamma(\rho_z)=(\min\{\sigma_1^2,
\sigma_2^2\})^{-1}$. 
Unless $\sigma_{1}$ and $\sigma_{2}$ are equal, the power off-set is
not monotonic over the interval $(-1,1)$.
Figure~\ref{fig:2b} shows the typical behavior of $\gamma(\rho_z)$.
\end{note}

 From Theorem~\ref{th:highSNR} we obtain the following corollary.
\begin{corollary}
\label{th:Th}
The prelog of the  Gaussian BC with noise-free feedback is 2 if $\rho_z
=-1$ or if $\rho_z=+1$ and $\sigma_1 \neq \sigma_2$; it is 1
otherwise: \begin{equation}\label{eq:rho-1}
  \varlimsup_{P \rightarrow \infty} \frac{C_{\BCFB,\Sigma}(P,\sigma_1^2, \sigma_2^2, \rho_z)}{\frac{1}{2} \log ( 1 + P )} \begin{cases} 2 & \rho_z=-1 \\
    2 & \rho_z =1 \textnormal{ and } \sigma_1^2 \neq \sigma_2^2 \\
    1 & \textnormal{otherwise}.\end{cases}
\end{equation}
\end{corollary}
\begin{note}\label{th:note}Our results for $\rho_z\in\{-1,1\}$ remain valid when the
  transmitter has only one-sided noise-free feedback, i.e., when the transmitter
  for example only observes the outputs $\{Y_{1,t}\}$ but
  not  $\{Y_{2,t}\}$. Indeed, for $\rho_z\in\{-1,1\}$, the
capacity regions with one-sided and two-sided noise-free feedback
coincide: when $\rho_z\in\{-1,1\}$ the transmitter can compute the
output it does not observe from the input and the output it does
observe.
\end{note}

Theorem~\ref{th:highSNR} and Corollary~\ref{th:Th} show that when
$\rho_z\in\{-1,1\}$ and the channel is not physically degraded,
noise-free feedback approximately doubles the high-SNR sum-rate
capacity. In Section~\ref{sec:Scheme} we present a coding scheme
achieving these gains. (When $\rho_z=-1$ also the Ozarow-Leung scheme
\cite{Ozarow:79,OzarowL:84} achieves such sum-rates
\cite{GastparW:08, GastparLW:10}.)
  \begin{figure}[tbp]
  \centering 
  \psfrag{t}[t][t]{\footnotesize $10^2$}
  \psfrag{f}[t][t]{\footnotesize $10^4$}
  \psfrag{o}[t][t]{\footnotesize$10^1$}
  \psfrag{tt}[t][t]{\footnotesize$10^3$}
    \psfrag{z}[t][t]{}
 \psfrag{2}[r][r]{\footnotesize$2$}
 \psfrag{1.8}[r][r]{\footnotesize$1.8$}
 \psfrag{1.6}[r][r]{\footnotesize$1.6$}
 \psfrag{1.4}[r][r]{\footnotesize$1.4$}
 \psfrag{1.2}[r][r]{\footnotesize$1.2$}
 \psfrag{1}[r][r]{\footnotesize$1$}
  \psfrag{0.6}[r][r]{}
 \psfrag{0.4}[r][r]{}
 \psfrag{0.2}[r][r]{}
 \psfrag{0}[r][r]{}
 \psfrag{0.8}[r][r]{\footnotesize$0.8$}
\psfrag{Rsum}[b][b]{\footnotesize$\frac{R_{\Sigma}(P)}{1/2\log(1+P)}$}
\psfrag{power}[t][t]{}
\psfrag{P}{$P$}
   \includegraphics[width=\columnwidth]{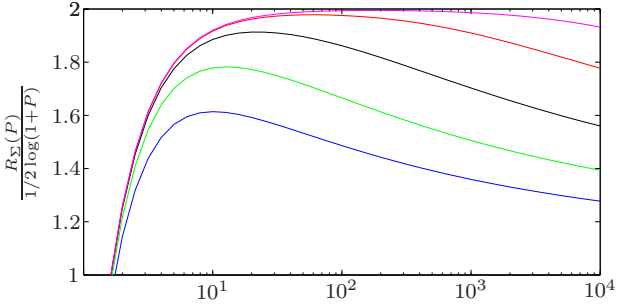}
     \caption{The sum-rate $R_\Sigma(P)$ achieved by the scheme in Section~\ref{sec:Scheme} normalized by $\frac{1}{2}\log(1+\frac{P}{\sigma_1^2})$ is plotted as a function of the power $P>0$. The noise variances $\sigma_1^2=\sigma_2^2=1$ are fixed, and the different curves correspond (in increasing order) to correlation coefficients $\rho_z =-0.85, -0.95, -0.99, -0.999 -0.9999$.} 
   \label{fig:4}
 \end{figure}
 Figure~\ref{fig:4} depicts the sum-rate achieved by the scheme of
 Section~\ref{sec:Scheme} as a function of the transmit power $P$, for
 various values of the correlation $\rho_z$.  It shows that for large
 powers $P$ (i.e., $P\geq 100$), noise-free feedback can nearly double
 the sum-rate capacity not only when the correlation $\rho_{z}$ is
 exactly $-1$, but also when $\rho_z$ is sufficiently close to -1,
 (i.e., when $\rho_z=-1 +\epsilon$ for sufficiently small $\epsilon
 >0$ depending on the power $P$). The same observation can be made
 when $\rho_z=1-\epsilon$ if $\sigma_1^2 \neq \sigma_2^2$.
 
 Theorem~\ref{thm:asympt} and Corollary~\ref{cor:1} ahead explore the
 relationship between the power $P$ and the correlation $\rho_{z}$
 that are required for noise-free feedback to roughly double the
 sum-rate capacity.  Since the required correlation depends on the
 transmit power $P$, we make the dependence explicit and denote the
 correlation by $\rho_{z}(P)$. Theorem~\ref{thm:asympt} and
 Corollary~\ref{cor:1} thus characterize the \emph{generalized prelog}
 where the channel parameters (here the noise correlation $\rho_z$)
 vary with the power~$P$.

 Let the noise variances $\sigma_1^{2}, \sigma_2^{2}>0$ be fixed. For
 every functional dependence $\rho_{z}(P)$ of the correlation
 coefficient on the power $P$, define
 \begin{IEEEeqnarray}{rCl}
\zminus&\triangleq &\varlimsup_{P\rightarrow \infty} \frac{-\log(1+\rho_z(P))}{\log(P)},\label{eq:conda}\\
\zplus& \triangleq&\varlimsup_{P\rightarrow \infty} \frac{-\log(1-\rho_z(P))}{\log(P)},\label{eq:condb}
\end{IEEEeqnarray}
where 
$-\log {0} \triangleq \infty$. {Notice that $\zminus>0$ only if $\varliminf_{P\to \infty} \rho_z(P)=-1$, and  $\zplus>0$ only if $\varlimsup_{P\to \infty} \rho_z(P)=1$.}

\begin{theorem}[Generalized Prelog with Noise-Free Feedback]\label{thm:asympt} 
 The generalized prelog depends on whether or not the noise variances 
 are equal.  If $\sigma_1^2 = \sigma_2^2$, then the generalized
 prelog is
\begin{equation}\label{eq:same}
\varlimsup_{P\rightarrow \infty}
\frac{C_{\BCFB,\Sigma}(P,\sigma_1^2, \sigma_2^2, \rho_z(P))}{\frac{1}{2}\log(1+P)}= \min\left\{1+\zminus,2\right\}
\end{equation}
and if $\sigma_1^2\neq \sigma_2^2$, then the generalized prelog  is
\begin{IEEEeqnarray}{rCl}\label{eq:alb}
\lefteqn{\varlimsup_{P\rightarrow \infty}
\frac{C_{\BCFB,\Sigma}(P,\sigma_1^2, \sigma_2^2, \rho_z(P))}{\frac{1}{2}\log(1+P)}} \qquad\qquad\qquad \nonumber \\ & =& \min\left\{1+ \max\Big\{\zminus,\zplus\Big\},2\right\}.\IEEEeqnarraynumspace
\end{IEEEeqnarray}
\end{theorem}
\begin{proof}
See Appendix~\ref{sec:pt2}.
\end{proof}

\begin{corollary}\label{cor:1}
Let $\rho_z(P)$ be of the form
\[
\rho_z(P) = \pm \left(1- \frac{\epsilon(P)}{P^{\zeta}} \right), \qquad \zeta \in[0,1]
\] 
where
\[
\lim_{P \to \infty} \frac{\log\bigl(\epsilon(P)\bigr)}{\log(P)} = 0.
\]
Unless $\sigma_1^2 = \sigma_2^2$ and $\lim_{P\to \infty} \rho_z(P)= 1$,
\begin{equation}
\varlimsup_{P\rightarrow \infty}
\frac{C_{\BCFB,\Sigma}(P,\sigma_1^2, \sigma_2^2, \rho_z(P))}{\frac{1}{2}\log(1+P)}
=1+\zeta.
\end{equation}
\end{corollary}

The above results on the dramatic capacity gains afforded by feedback
were predicated on the feedback being noise-free. Otherwise, as the
next theorem shows, the gains are more moderate.
\begin{theorem}
\label{th:NoisyRes}
  Irrespective of the correlation $\rho_z\in[-1,1]$, if the feedback
  is noisy then the prelog is one:
\begin{equation}\label{eq:noisy}
\varlimsup_{P\rightarrow \infty} \frac{
  C_{\BC\textnormal{Noisy},\Sigma}(P, \sigma_1^2, \sigma_2^2, \rho_z, \sigma_{W1}^2,\sigma_{W2}^2)} {\frac{1}{2} \log \left(1+P\right)} =
  1.
\end{equation}
\end{theorem}
\begin{proof}See Section~\ref{sec:PNoisyFB}.
\end{proof}

Thus, if the feedback is noisy, then the prelog equals 1 also when the
noise correlation $\rho_z$ is $\pm 1$. This result assumes that the
feedback-noise variances $\sigma_{W1}^2, \sigma_{W2}^2>0$ are
fixed. If instead they tend to 0 as the power $P\to\infty$, then for
$\rho_z\in\{-1,1\}$ the (generalized) prelog may be larger than 1,
depending on the speed of convergence of the limits $\sigma_{W1}^2,
\sigma_{W2}^2 \to 0$. The following note examines the generalized
prelog when the feedback-noise variances tend to 0 more slowly than
$P^{-\xi}$ for any
$\xi>0$. 
\begin{note}\label{rem:noisyfb}
Theorem~\ref{th:NoisyRes} remains valid if the feedback-noise variances $\sigma_{W1}^2,\sigma_{W2}^2$ tend to 0 as the power $P\to\infty$, if the convergence is slower than $P^{-\xi}$ for all $\xi>0$. More precisely, if $\sigma_{W1}^2, \sigma_{W2}^2$ depend on $P$ in a way that 
\begin{equation*}
\varlimsup_{ P\to \infty} \frac{-\log\left(\sigma_{W\nu}^2\right)}{\log(P)} \leq 0, \qquad \nu\in\{1,2\}, 
\end{equation*}
then the  prelog with noisy feedback is 1, irrespective of the noise correlation $\rho_z\in[-1,1]$. 
\end{note}
\begin{proof}
See Appendix~\ref{sec:Noisylim}.
\end{proof}


\subsection{A Coding Scheme for Noise-Free Feedback}\label{sec:proof-awgn-bc}
\label{sec:Scheme}

We present a ``successive noise cancellation'' coding strategy for the
Gaussian BC with noise-free feedback.  The scheme achieves the desired
rates in Theorems~\ref{th:highSNR} and \ref{thm:asympt}; see
Corollaries~\ref{th:rates}--\ref{lem:K20} and the proof of
Theorem~\ref{thm:asympt} in Appendix~\ref{sec:pt2}.

Our scheme is similar in flavor to the schemes proposed by Cover and
Pombra \cite{CoverP:89} for (non-white) Gaussian point-to-point
channels with noise-free feedback, by Lapidoth \& Wigger
 and  Khisti \& Lapidoth for the two-user Gaussian MAC with
noisy feedback \cite{LapidothW:10} or with intermittent feedback and side-information \cite{KhistiL:13}, and by Lapidoth, Steinberg, and Wigger
\cite{LapidothSW:10} for the two-user Gaussian BC with one-sided
noise-free feedback.

Before describing our scheme in Subsections~\ref{sec:BCscheme} and
\ref{sec:SchemeBCgen} ahead, we first motivate it by sketching a
simple scheme for the Gaussian point-to-point channel with noise-free
feedback (Subsection~\ref{sec:motscheme}) and a scheme for the
two-user Gaussian BC with noise-free feedback when $\rho_z$ is $\pm 1$
(Subsection~\ref{sec:schemeBC_1}).

\subsubsection{Motivation I: ``Successive Noise Cancellation'' scheme
  for the Gaussian point-to-point channel}
\label{sec:motscheme}
Consider $\eta$ transmissions over the standard memoryless Gaussian point-to-point  channel
\begin{equation}
Y_t = x_t +Z_t,
\end{equation}
where $x_t$ and $Y_t$ denote the time-$t$ input and output, $\{Z_t\}$ is a sequence of IID zero-mean, variance $\sigma^2>0$ Gaussian random variables, and the inputs $\{x_t\}$ are subject to an expected average block-power constraint $P$. The transmitter is assumed to have access to noise-free feedback. 

In the first channel use, the transmitted symbol is a unit-variance
information-carrying symbol 
$\Xi$. In the subsequent $\eta-1$
channel uses, the transmitted symbols are scaled versions of the
preceding noise symbols (which are known thanks to the feedback):
\begin{IEEEeqnarray}{rCl}
X_{1} & = & \sqrt{P} \, \Xi\\
X_{2}  & = & \sqrt{\frac{P}{\sigma^2}} \, Z_{1}\\
 \vdots \quad & = & \quad \vdots \nonumber \\
 X_{\eta} & = & \sqrt{\frac{P}{\sigma^2}} \, Z_{\eta -1}.
\end{IEEEeqnarray}
Consider now a (suboptimal) receiver that replaces the $\eta$ channel
outputs $Y_{1}, \ldots, Y_{\eta}$ with the single random variable~$I$,
where
\begin{IEEEeqnarray}{rCl}
I & = & \sum_{\ell=1}^\eta
\left(-\sqrt{\frac{P}{\sigma^2}}\right)^{\eta-\ell}
Y_{\ell} \label{eq:Ii} \\
& = & \sqrt{P}  \sqrt{\frac{P}{\sigma^2}}^{\eta-1} \Xi + Z_{\eta}. \label{eq:Isu}
\IEEEeqnarraynumspace
\end{IEEEeqnarray}
From~\eqref{eq:Isu} we see that the noise samples $Z_1, Z_2, \ldots,
Z_{\eta-1}$ have all been canceled out, and the only remaining noise
sample is $Z_{\eta}$. The channel from $\Xi$ to $I$ is a Gaussian
channel. Each use of this channel requires $\eta$ transmissions on the
original channel, so this scheme allows us to attain the rate
\begin{equation}\label{eq:rates}
  \frac{1}{2\eta} \log\left( 1 +  \left( \frac{P}{\sigma^{2}}\right)^{\eta} \right).
\end{equation}
  We obtain the largest achievable rate by choosing $\eta=1$. However, when we extend the scheme to the Gaussian BC, we will be interested in the limit $\eta \to \infty$. In this limiting case, the described scheme achieves any rate $R>0$ that satisfies
\begin{equation}
R< \frac{1}{2} \log^+\left(  \frac{P}{\sigma^{2}} \right).
\end{equation}
Thus, even though  for $\eta\to \infty$ the described scheme does not achieve capacity, it nevertheless achieves prelog 1 and the gap to capacity vanishes as $P$ tends to infinity.

The lesson from this example is that for high-SNR optimality it
suffices to send ``most of the time'' past noise samples,
rather than information symbols.


\subsubsection{Motivation II: ``Successive Noise Cancellation'' scheme for the Gaussian BC when $\rho_z\in\{-1,1\}$}\label{sec:schemeBC_1}

Consider the two-receivers Gaussian BC with noise-free feedback and noise
correlation $\rho_z\in\{-1,1\}.$
In this case, 
\begin{equation}
\frac{Z_{1,t}}{\sigma_1} =  \rho_z \frac{Z_{2,t}}{\sigma_2} \qquad \textnormal{with probability }1,
\end{equation}
i.e., the noise samples at the two receivers are proportional.
Consequently, using the successive noise cancellation scheme, the
transmitter can simultaneously and asymptotically-optimally serve both receivers.  To
see how, let the transmitted symbols be
\begin{IEEEeqnarray}{rCl}
X_{1} & = &\gamma( \Xi_1 + \Xi_2 )\\
X_{2}  & = &\gamma\left(\rho_z  \sqrt{\frac{P}{\sigma_2^2}}\Xi_1 + \sqrt{\frac{P}{\sigma_1^2}}\Xi_2\right) +  \sqrt{\frac{P}{\sigma_1^2}} Z_{1,1}\\
X_{3} & = &   \sqrt{\frac{P}{\sigma_1^2}} Z_{1,2} \\
 \vdots \quad & = & \quad \vdots \nonumber \\
 X_{\eta} & = & \sqrt{\frac{P}{\sigma_1^2}} Z_{1,\eta -1}
\end{IEEEeqnarray}
where $\gamma>0$ is a scaling factor that ensures that the sum of the powers of the first two inputs $X_1$ and $X_2$ does not exceed $2P$, and
where now we have two information-carrying symbols: $\Xi_1$ is
intended for Receiver~1 and $\Xi_2$ is intended for Receiver~2.
We transmit the two information symbols $\Xi_1$ and $\Xi_2$
along the two signaling directions, 
\begin{equation}
\vect{u}_1 = \trans{\begin{pmatrix}1 &  \rho_z \sqrt{P/\sigma_2^2} &0  & \ldots & 0\end{pmatrix} } 
\end{equation} 
and 
\begin{equation}
\vect{u}_2 =  \trans{\begin{pmatrix}1& \sqrt{P/\sigma_1^2} &0  & \ldots & 0\end{pmatrix} } ,
\end{equation}
which are different whenever the BC is not physically degraded, i.e.,
whenever $\rho_z \notin\big\{\frac{\sigma_1}{\sigma_2}, \frac{\sigma_2}{\sigma_1}\}$.

Receiver 1 uses $Y_{1,1}, \ldots, Y_{1,\eta}$ to compute $I_{1}$, where
\begin{IEEEeqnarray}{rCl}\label{eq:Ii1}
I_1 & = & \sum_{\ell=1}^\eta \left( - \sqrt{\frac{P}{\sigma_1^2}}\right)^{\eta-\ell} Y_{1,\ell}\label{eq:Isu1}\nonumber \\ 
&=  &\gamma \left( - \sqrt{\frac{P}{\sigma_1^2}}\right)^{\eta-1}\left( 1 - \rho_z \frac{\sigma_1}{\sigma_2} \right) \Xi_1 + Z_{1,\eta},    \IEEEeqnarraynumspace
\end{IEEEeqnarray}
and Receiver 2  uses $Y_{2,1}, \ldots, Y_{2,\eta}$ to  compute $I_{2}$, where
\begin{IEEEeqnarray}{rCl}\label{eq:Ii2}
I_2 & = & \sum_{\ell=1}^\eta \left(- \rho_z \sqrt{\frac{P}{\sigma_2^2}}\right) ^{\eta-\ell} Y_{2,\ell} \nonumber \\
\label{eq:Isu2}&=&  \gamma\left( -\sqrt{\frac{P}{\sigma_2^2}}\right)^{\eta-1} \left( 1 - \rho_z \frac{\sigma_2}{\sigma_1}\right) \Xi_2 + Z_{2,\eta}.   \IEEEeqnarraynumspace
\end{IEEEeqnarray}
In other words, each receiver projects its observed outputs onto a particular receive beam-forming vector: 
\begin{equation}
\vect{v}_1 = \trans{\begin{pmatrix} \big(-\sqrt{{P}/{\sigma_1^2}}\big)^{\eta-1} & \big(-\sqrt{{P}/{\sigma_1^2}}\big)^{\eta-2}& \ldots &
 1\end{pmatrix}} 
\end{equation}
at Receiver~1 and 
\begin{equation}
\vect{v}_2 = \trans{\begin{pmatrix} \big(-\rho_z \sqrt{{P}/{\sigma_2^2}}\big)^{\eta-1} & \big(-\rho_z\sqrt{{P}/{\sigma_2^2}}\big)^{\eta-2}& \ldots &
 1\end{pmatrix}}
\end{equation}
at Receiver~2. These beam-forming vectors are different whenever the Gaussian BC is not physically degraded.

We see from \eqref{eq:Isu1} and \eqref{eq:Isu2} that the noise samples
$Z_{1,1}, \ldots, Z_{1,\eta-1}$ and $Z_{2,1}, \ldots, Z_{2,\eta-1}$
are completely canceled and do not influence $I_{1}$ and
$I_{2}$. Only the last noise samples $Z_{1,\eta}$ and $Z_{2,\eta}$
remain. Moreover, also the undesired information symbol is canceled
out: $\Xi_2$ does not influence $I_1$, and $\Xi_1$ does not influence $I_2$.
The channel from $\Xi_{1}$ to $I_{1}$ and the channel from $\Xi_{2}$
to $I_{2}$ are parallel Gaussian channels (with dependent
noises). We thus obtain from~\eqref{eq:Isu1}
and~\eqref{eq:Isu2} 
the achievability of the rates 
\begin{IEEEeqnarray}{rCl}\label{eq:rates12}
R_1 &=& \frac{1}{2\eta} \log \left( 1+ \frac{\gamma^2}{P} \left( \frac{P}{\sigma_1^2}\right)^{\eta} \left( 1 - \rho_z \frac{\sigma_1}{\sigma_2} \right)^2  \right) \\
R_2& = & \frac{1}{2\eta} \log \left( 1+ \frac{\gamma^2}{P} \left( \frac{P}{\sigma_2^2}\right)^{\eta} \left( 1 - \rho_z \frac{\sigma_2}{\sigma_1} \right)^2  \right) .
\end{IEEEeqnarray}
Notice that $\gamma$ is nonzero and does not depend on
$\eta$. Consequently, when the BC is not physically degraded,
\eqref{eq:pd}, and when $\eta$ tends to infinity, the described scheme
achieves all rate pairs $(R_1,R_2)$ that satisfy (see also
Corollary~\ref{lem:K20} ahead)
\begin{IEEEeqnarray}{rCl}
R_1 &<& \frac{1}{2} \log^+ \left(\frac{P}{\sigma_1^2}\right)  \\
R_2& <& \frac{1}{2} \log^+ \left( \frac{P}{\sigma_2^2} \right) .
\end{IEEEeqnarray}

\subsubsection{A General Scheme}\label{sec:BCscheme}

The ``Successive Noise Cancellation'' schemes can be extended to the
general Gaussian BC with noise-free feedback. The idea is to use the
feedback to transform each block of $\eta$ uses of the original scalar
BC into a single use of a new MISO BC, which can be viewed as a BC with two
transmit antennas and a single receive antenna for each
receiver: the new BC's input is the vector
$\trans{(\Xi_1,\Xi_2)}\in\Reals^2$, and its two scalar outputs are
$I_1\in\Reals$ at Receiver~1 and $I_2\in\Reals$ at Receiver~2. We then
code over this new BC ignoring the feedback. Scaling by $\eta^{-1}$
any rate pair that is achievable on the new BC will yield a pair that
is achievable on the original BC with feedback.

We next describe how to transform a block of $\eta$ uses of the
original scalar BC with feedback into a single use of the MISO BC. For
simplicity, we restrict attention to the first block; the procedure
for the subsequent blocks is analogous. The key parameters are the
signaling vectors $\vect{u}_1$ and $\vect{u}_2$, the coefficients
according to which past noise symbols are retransmitted, which we
collect into the strictly lower-triangular matrices $\mat{B}_1,
\mat{B}_2$, and the receivers' beam-forming vectors $\vect{v}_1$ and
$\vect{v}_2$.  (In the following Subsection~\ref{sec:SchemeBCgen}, we
present a specific choice for these parameters.)

\newcommand{\bftZ}{\vect{{Z}}}
\newcommand{\bftY}{\vect{{Y}}}

We use the notation $\bfX\triangleq (X_1,\ldots, X_\eta)$,
$\bfY_1\triangleq (Y_{1,1},\ldots, Y_{1,\eta})$, and $\bfY_2\triangleq
(Y_{2,1},\ldots, Y_{2,\eta})$. %
The transmitter produces the $\eta$-length input vector
$\bfX$ 
based on the information carrying symbols $\Xi_1$ and $\Xi_2$ and on the
feedback signals it receives:\footnote{The transmitter can compute all
  the past noise symbols because, through the feedback, it learns the
  past channel outputs and because it also knows the past channel
  inputs.}
\begin{equation}\label{newinputs}
\bfX = \Xi_{1} \vect{u}_1+ \Xi_{2} \vect{u}_2 +  \mat{B}_1  \bftZ_{1}  +  \mat{B}_2 \bftZ_{2},
\end{equation}
where $\bfZ_1\triangleq (Z_{1,1},\ldots, Z_{1,\eta})$ and
$\bfZ_2\triangleq (Z_{2,1},\ldots, Z_{2,\eta})$, and where the
assumption that $\mat{B}_1$ and $\mat{B}_2$ are lower-triangular
guarantees that $\Xi_1,\Xi_2$, and the feedback signals suffice to
compute~$\bfX$.

The inputs to the original channel satisfy the average block-power
constraint \eqref{eq:power} whenever the information carrying symbols
$\Xi_1$ and $\Xi_2$ are independent and satisfy $\E{\Xi_1^2},
\E{\Xi_2^2}\leq 1$ and
when 
\begin{IEEEeqnarray}{rCl}\label{eq:powerblock}
\lefteqn{
 \|\vect{u}_1\|^2 +  \|\vect{u}_2\|^2  + \trace{ \mat{B}_1\trans{\mat{B}}_1}\sigma_1^2+ \trace{ \mat{B}_2\trans{\mat{B}}_2}\sigma_2^2} \nonumber\\ & & \hspace{3.6cm} + 2 \trace{\mat{B}_1 \trans{\mat{B}}_2} \rho_z \sigma_1\sigma_2 \leq \eta P. \hspace{0.6cm}
\end{IEEEeqnarray}

Receiver~$1$ observes
\begin{IEEEeqnarray}{rCl}\label{eq:Yout1}
\bfY_{1}
& = & 
\Xi_{1} \vect{u}_1 +  \Xi_{2} \vect{u}_2 + \left( \mat{B}_{1}+\mat{I} \right)\bftZ_{1} + \mat{B}_{2} \bftZ_{2}, \IEEEeqnarraynumspace
\end{IEEEeqnarray}
and computes 
\begin{IEEEeqnarray}{rCl}\label{eq:I1}
I_{1}&\triangleq& \trans{\vect{v}}_{1} \bfY_{1}.
\IEEEeqnarraynumspace
\end{IEEEeqnarray}
 Receiver~$2$ observes
\begin{IEEEeqnarray}{rCl} 
 \bfY_{2}& =& \Xi_{1} \vect{u}_1 + \Xi_{2} \vect{u}_2 + \left( \mat{B}_{2}+\mat{I} \right)\bftZ_{2} + \mat{B}_{1} \bftZ_{1}, \IEEEeqnarraynumspace
\end{IEEEeqnarray}
and computes
\begin{IEEEeqnarray}{rCl}\label{eq:I2}
I_{2}&\triangleq& \trans{\vect{v}}_{2} \bfY_{2}. 
\IEEEeqnarraynumspace
\end{IEEEeqnarray}

\subsubsection{Choice of Parameters and Achievable Rates}\label{sec:SchemeBCgen}
Given $\eta$, we describe a choice of the parameters $\mat{B}_1, \mat{B}_2$, $\vect{u}_1, \vect{u}_2$, $\vect{v}_1, \vect{v}_2$. 

Choose
$q>0$ and $\delta\notin\{-1,0\}$ to satisfy Equation~\eqref{eq:bpower} (shown at the top of the next page), and
\begin{figure*}
\begin{IEEEeqnarray}{rCl}\label{eq:bpower}
q^2 (\sigma_1^2 + \delta^4 \sigma_2^2 -2 \delta^2 \rho_z \sigma_1\sigma_2) + q^4 (1+\delta)^2 \delta^2 (\sigma_1^2 + \delta^2 \sigma_2^2 + 2 \delta \rho_z \sigma_1 \sigma_2)  \leq P.
\end{IEEEeqnarray}
\hrule
\end{figure*}
define
\begin{IEEEeqnarray}{rCl}
a_1 & \triangleq& q \label{eq:a1}\\
a_2 & \triangleq & -\delta^2 q \label{eq:a2}\\
b_1& \triangleq & -\delta (1+\delta)q^2\label{eq:b1}\\
b_2& \triangleq & -\delta^2 (1+\delta)q^2.\label{eq:b2}
\end{IEEEeqnarray}  
Choose the $\eta\times\eta$ matrices $\mat{B}_1$ and $\mat{B}_2$ to be
Toeplitz with non-zero entries only on the first and second diagonals
below the main diagonal:
\begin{equation}\label{eq:Bk}
\mat{B_k}= \begin{pmatrix}
0 & 0 & 0 & \cdots & 0 & 0 & 0\\
a_k & 0 &  0& \cdots & 0  & 0&0 \\
b_k& a_k& 0 & \cdots & 0 & 0 &0  \\
0 & b_k & a_k &0 &\cdots & \vdots& \vdots\\
\vdots& \ddots& \ddots& \ddots& \ddots & \vdots& \vdots
\\
0 & \cdots &0 & b_k & a_k & 0 &0\\
0 & \cdots &\cdots & 0&b_k & a_k & 0
\end{pmatrix},
\end{equation} 
and  choose the $\eta$-dimensional vectors
\begin{subequations}\label{eq:u12}
\begin{IEEEeqnarray}{rCl}
\vect{u}_1&=&\sqrt{\frac{P}{2 + 2 \frac{b_1^2}{a_1^2}}}\trans{\begin{pmatrix}1& \frac{b_1}{a_1} & 0&\ldots& 0\end{pmatrix}}\\
\vect{u}_2&=&\sqrt{\frac{P}{2 + 2 \frac{b_2^2}{a_2^2}}}\trans{\begin{pmatrix}1& \frac{b_2}{a_2}& 0&\ldots& 0\end{pmatrix}}
\end{IEEEeqnarray}
\end{subequations}
and 
\begin{subequations}\label{eq:v12}
\begin{IEEEeqnarray}{rCl}
\vect{v}_1& =& \trans{\begin{pmatrix} \left( -\frac{b_2}{a_2}\right)^{\eta-1}&\left( -\frac{b_2}{a_2}\right)^{\eta-2}  &\ldots & -\frac{b_2}{a_2}& 1 \end{pmatrix}}\\
\vect{v}_2& =& \trans{\begin{pmatrix} \left( -\frac{b_1}{a_1}\right)^{\eta-1}&\left( -\frac{b_1}{a_1}\right)^{\eta-2}  &\ldots & -\frac{b_1}{a_1}& 1 \end{pmatrix}}.\IEEEeqnarraynumspace\label{eq:v2}
\end{IEEEeqnarray} 
\end{subequations}
By \eqref{eq:bpower}--\eqref{eq:u12}, 
this choice satisfies the power constraint
\eqref{eq:powerblock}. Moreover, the vector $\vect{v}_1$ is orthogonal
to the first $\eta-2$ columns of the matrices $(\mat{B}_1 + \mat{I})$
and $\mat{B}_2$, and to the vector $\vect{u}_2$, but not to
$\vect{u}_1$. Similarly, $\vect{v}_2$ is orthogonal to the first
$\eta-2$ columns of the matrices $\mat{B}_1$ and $(\mat{B}_2 +
\mat{I})$, and to the vector $\vect{u}_1$, but not to $\vect{u}_2$.
Therefore, the noise samples $Z_{1, 1}, \ldots, Z_{1,\eta -2}$ and
$Z_{2, 1}, \ldots, Z_{2,\eta -2}$ are completely canceled out when
forming the ``new outputs'' in \eqref{eq:I1} and only the noise
samples $Z_{1,\eta-1}, Z_{1,\eta}, Z_{2,\eta-1}, Z_{2,\eta}$ remain.
Moreover, the ``interference'' $\Xi_{2}$ is canceled in $I_{1}$ and the
``interference'' $\Xi_{1}$ is canceled in $I_{2}$. In fact, 
\begin{subequations}\label{eq:I}
\begin{IEEEeqnarray}{rCl}
I_{1} & = & \sqrt{\frac{P}{2 + 2 \frac{b_1^2}{a_1^2}}}  \left( -\frac{b_2}{a_2}\right)^{\eta-1} \left( 1- \frac{a_2}{b_2} \frac{b_1}{a_1}\right) \Xi_{1,i}\nonumber\\
 &&  + \left( -\frac{b_2}{a_2} + a_1\right) Z_{1,\eta-1} + Z_{1,\eta} + a_2 Z_{2,\eta-1}\IEEEeqnarraynumspace
 \end{IEEEeqnarray}
 and 
\begin{IEEEeqnarray}{rCl}
I_{2} & = & \sqrt{\frac{P}{2 + 2 \frac{b_2^2}{a_2^2}}}  \left( -\frac{b_1}{a_1}\right)^{\eta-1} \left( 1- \frac{a_1}{b_1} \frac{b_2}{a_2}\right) \Xi_{2,i} \nonumber \\
&&+ \left( -\frac{b_1}{a_1} + a_2\right) Z_{2,\eta-1} + Z_{2,\eta} + a_1 Z_{1,\eta-1}.\IEEEeqnarraynumspace
\end{IEEEeqnarray}
\end{subequations} 

Over the original Gaussian BC with feedback we can achieve the
scaled-by-$\eta^{-1}$ capacity of the new MISO BC \eqref{eq:I}, and we
thus have the following proposition.
\begin{proposition}\label{prop:achrates}
The noise-free feedback scheme of Section~\ref{sec:BCscheme}
with the choice of parameters presented here in
Section~\ref{sec:SchemeBCgen}, achieves all rate pairs $(R_1, R_2)$
for which
\begin{subequations}\label{eq:r12}
\begin{IEEEeqnarray}{rCl}
R_1 & \leq  & \frac{1}{2\eta} \log\left( 1+ \frac{\frac{P(1+\delta)^2}{2 + 2 q^2\delta^2(1+\delta)^2}\left(  q^2(1+\delta)^2\right)^{\eta-1}}{  (q^2\delta^2+1)\sigma_1^2 +  q^2 \delta^4\sigma_2^2} \right) \\
R_2 & \leq & \frac{1}{2\eta} \log\left( 1+ \frac{\frac{P(1+1/\delta^2)^2}{2 + 2 q^2(1+\delta)^2}\left(  q^2\delta^2(1+\delta)^2\right)^{\eta-1}}{  (q^2\delta^2+1)\sigma_2^2 +  q^2 \sigma_1^2} \right)\IEEEeqnarraynumspace
\end{IEEEeqnarray}
\end{subequations}
simultaneously hold for some real numbers $\delta \notin\{-1, 0\}$ and
$q$ such that \eqref{eq:bpower} holds.
\end{proposition}

The choice of parameters in~\eqref{eq:a1}--\eqref{eq:v12} that leads
to Proposition~\ref{prop:achrates} is, in general, sub-optimal; better
choices can be found in~\cite{GastparLSW:11}.
However, whenever the BC is not physically degraded, the rates in
Proposition~\ref{prop:achrates} achieve the asymptotic high-SNR
sum-rate capacity with noise-free feedback (Theorem~\ref{th:highSNR});
see Corollaries~\ref{cor:highSNR} and \ref{lem:K20} ahead. Moreover,
they also achieve the generalized prelog in
Theorem~\ref{thm:asympt}. In fact, Corollary~\ref{th:rates} ahead
suffices to prove the achievability of Theorem~\ref{thm:asympt}, as is
shown in Appendix~\ref{sec:pt2}.


\subsubsection{High-SNR Performance}\label{sec:highSNR}

By the following lemma, $\eta=1$ maximizes the constraints in \eqref{eq:r12} for small powers $P$ and $\eta \to\infty$ maximizes them for large powers $P$.  
\begin{lemma}\label{lem:opteta}
Let $\xi, \zeta$ be positive real numbers. 
If $1+\zeta \geq \xi$, then the  mapping $\eta \in\mathbb{Z}^+ \mapsto \frac{1}{2\eta} \log( 1+  \xi^{\eta-1} \zeta)$, has its maximum at $\eta=1$; otherwise it has its supremum at $\eta\to \infty$.
\end{lemma} 
\begin{proof}
See Appendix~\ref{app:lemma}.
\end{proof}

Letting $\eta\to \infty$, we  obtain the following corollary to Proposition~\ref{prop:achrates}.
\begin{corollary}\label{th:rates}
All nonnegative rate-pairs $(R_1,R_2)$ that satisfy
\begin{subequations}\label{eq:ratesgen}
\begin{IEEEeqnarray}{rCl}
R_1&<  & \frac{1}{2} \log^+ \left(q^2 (1+\delta)^2\right)\\
R_2& < & \frac{1}{2} \log^+ \left(q^2 \delta^2(1+\delta)^2\right),
\end{IEEEeqnarray}
\end{subequations}
for some real numbers $\delta \notin\{-1,0\}$ and $q$ such that \eqref{eq:bpower} holds, are achievable over the Gaussian BC with noise-free feedback.
\end{corollary}

From Corollary~\ref{th:rates} with an appropriate choice of the
parameters $\delta\notin\{-1,0\}$ and $q$ we further obtain:
\begin{corollary}\label{cor:highSNR}
If $\rho_z\in(-1,1)$, then for every $\epsilon\in(0,1)$ there exists a positive real number $P_0(\epsilon, \sigma_1^2, \sigma_2^2, \rho_z)$ such that the sum-rate 
\begin{IEEEeqnarray}{rCl}
R_{1}+R_2= \frac{1}{2} \log^+ \left( \frac{(1-\epsilon)P(\sigma_1^2 +\sigma_2^2 - 2\rho_z \sigma_1 \sigma_2)}{\sigma_1^2 \sigma_2^2 (1-\rho_z^2)}\right)\IEEEeqnarraynumspace
\end{IEEEeqnarray}
is achievable over the Gaussian BC with noise-free feedback when the
allowed power $P$ exceeds $P_0(\epsilon, \sigma_1^2, \sigma_2^2, \rho_z)$. 
\end{corollary}
\begin{proof}
By choosing
\begin{subequations}\label{eq:choice0}
\begin{IEEEeqnarray}{rCl}
\delta & = &\frac{\sigma_1}{\sigma_2}\cdot \frac{ \sigma_1 -\rho_z \sigma_2}{\sigma_2 -\rho_z \sigma_1}\\
q&=  & \left( \frac{(1-\epsilon) P}{ \delta^2 (1+\delta)^2 (\sigma_1^2 + \delta ^2\sigma_2^2 + 2 \delta \rho_z \sigma_1 \sigma_2)}\right)^{1/4},
\end{IEEEeqnarray}
\end{subequations}
in Corollary~\ref{th:rates} and in power constraint \eqref{eq:bpower}. 
\end{proof}
\begin{corollary}\label{lem:K20}
If $\rho_z\in\{-1,1\}$, and the channel is not physically degraded, i.e., $\rho_z\notin\left\{ \frac{\sigma_1}{\sigma_2}, \frac{\sigma_2}{\sigma_1}\right\}$, then all nonnegative rate-pairs $(R_1,R_2)$ that satisfy
\begin{subequations}
\begin{IEEEeqnarray}{rCl}
R_1 & <  & \frac{1}{2} \log^+\left(\frac{P}{\sigma_1^2}\right) \\
R_2 & < & \frac{1}{2} \log^+ \left( \frac{P}{\sigma_2^2}\right)
\end{IEEEeqnarray}
\end{subequations}
are achievable over the Gaussian BC with noise-free feedback.
\end{corollary}
\begin{proof} 
Follows from Corollary~\ref{th:rates} by choosing
%
\begin{subequations}\label{eq:choice1}
\begin{IEEEeqnarray}{rCl}
\delta & = & -\rho_z\frac{\sigma_1}{\sigma_2}\\
q&=  & \Bigg( \frac{P}{\sigma_1^2\big(1-\frac{\sigma_1}{\sigma_2}\rho_z\big)^2}\Bigg)^{1/2}
\end{IEEEeqnarray}
\end{subequations}
and verifying the power constraint~\eqref{eq:bpower}.
\end{proof}

%

\begin{note}
Specializing the rate constraints in \eqref{eq:r12} to  the choice in \eqref{eq:choice1} we conclude that  when $\rho_z\in\{-1,1\}$ all rate-pairs $(R_1,R_2)$ that satisfy
\begin{subequations}\label{eq:rateseta1}
\begin{IEEEeqnarray}{rCl}
R_1 & \leq  & \frac{1}{2\eta} \log\left( 1+ \left( \frac{P}{\sigma_1^2}\right)^{\eta-1} \left(  1 - \rho_z\frac{\sigma_1}{ \sigma_2}\right)^2 \frac{P/\sigma_1^2}{2 + 2 P/\sigma_2^2}\right)\nonumber \\ \\
R_2 & \leq & \frac{1}{2\eta} \log\left( 1+\left( \frac{P}{\sigma_2^2}\right)^{\eta-1} \left(  1 - \rho_z\frac{ \sigma_2}{\sigma_1}\right)^2 \frac{P/\sigma_2^2}{2 + 2 P/\sigma_1^2} \right)\nonumber \\\IEEEeqnarraynumspace
\end{IEEEeqnarray}
\end{subequations}
 for some positive integer $\eta$,
are achievable.
Consequently, for fixed $\eta$,  when $\rho_z\in\{-1,1\}$ and  the BC is not physically degraded,  our scheme achieves prelog $2\frac{\eta-1}{\eta}$. 
 Thus, $\eta=3$ suffices to increase the prelog compared to the non-feedback setup. Also,  when $\eta \to\infty$ the scheme can achieve prelog~2. 
\end{note}

\subsection{Proof of Theorem~\ref{th:NoisyRes} (Prelog with Noisy Feedback)}\label{sec:PNoisyFB}
The interesting part is the converse, which we prove using a
genie-argument inspired by~\cite{KimLapidothWeissman:06}. It is based
on the following three steps. 1.) We introduce a \emph{genie-aided
  Gaussian BC} without feedback and show that its sum-rate capacity upper
bounds the sum-rate capacity of the original Gaussian BC with noisy
feedback. 2.) We introduce a \emph{less noisy Gaussian BC} with neither
genie-information nor feedback and show that its sum-rate capacity
coincides with the sum-rate capacity of the genie-aided Gaussian BC. 3.)
We show that the prelog of the less noisy Gaussian BC equals 1,
irrespective of the noise variances
$\sigma_1^2,\sigma_2^2,\sigma_{W1}^2, \sigma_{W2}^2>0$ and the
correlation coefficient $\rho_z\in[-1,1]$.

We next elaborate on these steps starting with the first.  The
genie-aided Gaussian BC is defined as the original Gaussian BC \emph{without
  feedback}, but with a genie that prior to transmission reveals the
sequences $\{(Z_{1,t}+W_{1,t})\}_{t=1}^n$ and
$\{(Z_{2,t}+W_{2,t})\}_{t=1}^n$ to the transmitter and both
receivers. Notice that with this genie information, after each channel
use $t$, the transmitter can compute the missing feedback
outputs $V_{1,t}$ and $V_{2,t}$:
  \begin{IEEEeqnarray}{rCl}\label{eq:V1d}
V_{1,t}&=& X_t+(Z_{1,t}+W_{1,t}), \\ V_{2,t} &= &
  X_t+(Z_{2,t}+W_{2,t}).\label{eq:V2d} 
\end{IEEEeqnarray} 
Consequently, the sum-rate capacity of the genie-aided
Gaussian BC is at least as large as the sum-rate capacity of the original
Gaussian BC with feedback. 

We next elaborate on the second step. 
The less noisy Gaussian BC  is described by the channel law 
\begin{IEEEeqnarray}{rCl}
{Y}_{1,t}' & = & x_t +{Z}_{1,t}',\label{eq:Yt1}\\
{Y}_{2,t}' & = & x_t+{Z}_{2,t}',\label{eq:Yt2}
\end{IEEEeqnarray}
where the reduced noise samples ${Z}_{1,t}'$, and ${Z}_{2,t}'$ are defined as
 \begin{IEEEeqnarray}{rCl}
{Z}_{1,t}' & \triangleq  & Z_{1,t} - \E{Z_{1,t}|
  (Z_{1,t}+W_{1,t}), (Z_{2,t}+W_{2,t})}, \label{eq:tZ1}
  \IEEEeqnarraynumspace \\ 
{Z}_{2,t}' & \triangleq & Z_{2,t} - \E{Z_{2,t}| (Z_{1,t}+W_{1,t}), (Z_{2,t}+W_{2,t})}, \label{eq:tZ2}\IEEEeqnarraynumspace
\end{IEEEeqnarray}
and are of variances
\begin{IEEEeqnarray}{rCl}
\Var{{Z}_{1,t}'} & = &  \sigma_1^2 \frac{\sigma_{W1}^2 \sigma_2^2(1-\rho_z^2) +
  \sigma_{W1}^2
  \sigma_{W2}^2}{(\sigma_1^2+\sigma_{W1}^2)(\sigma_2^2+\sigma_{W2}^2)
  -\sigma_1^2\sigma_2^2\rho_z^2}, \IEEEeqnarraynumspace \label{eq:Var1}\\
\Var{{Z}_{2,t}'} & = &  \sigma_2^2 \frac{\sigma_{W2}^2 \sigma_1^2(1-\rho_z^2) +
  \sigma_{W2}^2
  \sigma_{W1}^2}{(\sigma_1^2+\sigma_{W1}^2)(\sigma_2^2+\sigma_{W2}^2)
  -\sigma_1^2\sigma_2^2\rho_z^2} .\IEEEeqnarraynumspace \label{eq:Var2}
\end{IEEEeqnarray}
By the following two observations, the sum-rate capacity of this less
noisy Gaussian BC coincides with the sum-rate capacity of the genie-aided
Gaussian BC.  The first is that the sum-rate capacity of the less noisy
Gaussian BC remains unchanged if prior to transmission a genie reveals the
sequences $\{Z_{1,t}+W_{1,t}\}$ and $\{Z_{2,t}+W_{2,t}\}$ to the
transmitter and both receivers. Indeed,
by~\eqref{eq:tZ1}--\eqref{eq:tZ2} and the Gaussianity of all involved
sequences the genie-information $\{Z_{1,t}+W_{1,t}\}$ and $\{
Z_{2,t}+W_{2,t}\}$ is independent of the reduced noise sequences
$\{{Z}_{1,t}',{Z}_{2,t}'\}$, and it thus plays only the role of common
randomness, which does not increase capacity.  The second observation
is that the sum-rate
capacity of the genie-aided Gaussian BC coincides with the sum-rate
capacity of the less noisy Gaussian BC, when in this latter case the
transmitter and both receivers additionally know the genie-information
$\{(Z_{1,t}+W_{1,t})\}$ and $\{(Z_{2,t}+W_{2,t})\}$.  Indeed, 
knowing the genie-information $\{(Z_{1,t}+W_{1,t})\}_{t=1}^n$
and $\{(Z_{2,t}+W_{2,t})\}_{t=1}^n$, the outputs ${Y}_{1,t}'$ and
${Y}_{2,t}'$ can be transformed into the outputs $Y_{1,t}$ and
$Y_{2,t}$, and vice versa.

We finally elaborate on the third step. The less noisy Gaussian BC is a
classical Gaussian BC with neither feedback nor genie-information, and its
sum-rate capacity is   \cite{Bergmans:74}
\begin{IEEEeqnarray}{rCl}\lefteqn{
C_{\textnormal{BCLessNoisy},\Sigma}\big(P,\sigma_1^2, \sigma_2^2, \rho_z,
\sigma_{W1}^2, \sigma_{W2}^2\big)}\quad \nonumber \\& =& \frac{1}{2} \log\left( 1+ \frac{ P
  }{\min\left\{ \Var{{Z}_{1,t}'}, \Var{{Z}_{2,t}'} \right\}
  }\right),\IEEEeqnarraynumspace
\label{eq:sumc}
\end{IEEEeqnarray}
where the variances $\Var{{Z}_{1,t}'}, \Var{{Z}_{2,t}'} $ are defined in \eqref{eq:Var1} and \eqref{eq:Var2}.
By \eqref{eq:Var1}--\eqref{eq:sumc} the prelog of the less noisy
Gaussian BC equals 1, irrespective of the noise variances $\sigma_1^2,
\sigma_2^2, \sigma_{W1}^2,\sigma_{W2}^2>0$ and the noise correlation
$\rho_z\in[-1,1]$. 
This concludes the third step, and thus our proof.

\section{$K$-User Broadcast Channel}\label{sec:KBC}
\subsection{Setup and Results}
We next extend the model of Section~\ref{sec:pfb} by allowing
the number of receivers $K$ to exceed two. We  assume
noise-free feedback.
For each $k \in \{1,\ldots, K\}$ we denote the message intended for
Receiver~$k$ by $M_k$, and we assume that it is uniformly distributed over
$\set{M}_k\triangleq\{1,\ldots, \lfloor 2^{nR_k}\rfloor\}$ and that
$M_1,\ldots, M_K$ are independent.
The time-$t$ symbol observed by Receiver~$k$ is
\begin{equation}\label{eq:BCK}
Y_{k,t} = x_{t}  +Z_{k,t}, \qquad t\in\{1,\ldots, n\}, 
\end{equation}
where $x_{t}$ is the time-$t$ transmitted symbol, and $Z_{k,t}$ is the
time-$t$ noise sample at Receiver~$k$. We assume that
$\{\trans{(Z_{1,t}, \ldots, Z_{K,t})}\}_{t=1}^n$ is a sequence of IID
centered Gaussian vectors of covariance matrix $\mat{K}_z$ and that
this sequence is independent of the messages $(M_1,\ldots, M_K)$. We denote the
variance of the noise at the $k$-th receiver by $\sigma_{k}^{2}$ and
the standard deviation by $\sigma_{k}$. We assume that the
standard deviations are all strictly positive
\begin{equation}
  \label{eq:amos100}
  \sigma_{k} > 0, \qquad k \in \{1, \ldots, K\}.
\end{equation}

Based on the messages and the feedback signals, the transmitter 
produces the time-$t$ channel input
\begin{equation*}
X_t=f_{\textnormal{K-BC},t}^{(n)}(M_1, \ldots, M_K, Y_1^{t-1}, \ldots, Y_K^{t-1})
\end{equation*}
using encoding functions of the form
\begin{equation}
f_{\textnormal{K-BC},t}^{(n)}\colon \set{M}_1 \times \cdots \times \set{M}_K \times \Reals^{K(t-1)} \rightarrow \Reals
\end{equation}
that are constrained to produce channel inputs $X_1,\ldots, X_n$
satisfying the expected average block-power
constraint~\eqref{eq:power}.

Based on its received sequence $Y_{k,1}, \ldots, Y_{k,n}$,
Receiver~$k$ forms the guess $\hat{M}_{k}$ of $M_{k}$.  We say that an
error occurred whenever at least one of the receivers errs, i.e., whenever
\begin{equation}
(M_1,\ldots, M_K) \neq (\hat{M}_1,\ldots, \hat{M}_K).
\end{equation}
Achievable rate-tuples, the capacity region, the sum-rate capacity,
and the prelog are defined as in the two-receiver case. We denote the
sum-rate capacity by $C_{\textnormal{K-BC},\Sigma}(P, \mat{K}_{z})$.

Our main result for this model is the prelog when 
\begin{equation}\label{eq:rankKZ}
\textnormal{rank}(\mat{K}_z)=1.
\end{equation} 
This case is the $K$-receivers analog of the two-receiver setting with
noise correlation $\pm 1$. In this case, the noise samples $Z_{1,t},
\ldots, Z_{K,t}$ are all multiples of each other, and we can rewrite
the channel law as:
\begin{equation}\label{eq:BCK1}
Y_{k,t} = X_{t}  +\rho_{1,k} \sigma_k {Z}_{1,t}, \qquad t\in\{1,\ldots, n\},
\end{equation}
where 
$\sigma_k>0$ (by~\eqref{eq:amos100}),  and $\rho_{1,k}$ denotes the correlation
coefficient between $Z_{1,t}$ and $Z_{k,t}$, which,
by~\eqref{eq:rankKZ}, is either $-1$ or
$1$:
\begin{equation}
  \label{eq:amos200}
  \rho_{1,k} \in \{-1,+1\}, \qquad k\in\{1,\ldots, K\}.
\end{equation}
Define
\begin{equation}\label{eq:alk}
\alpha_k \triangleq \rho_{1,k} \sigma_k, \quad k \in \{1, \ldots, K\},
\end{equation}
and note that by~\eqref{eq:amos100} and~\eqref{eq:amos200}
\begin{equation}\label{eq:alphazero}
\alpha_k\neq 0, \qquad k \in \{1, \ldots, K\}.
\end{equation}

As we shall see, when $\mat{K}_z$ is of rank~$1$, the prelog depends
on the number $n_\alpha$ of $\alpha_k$'s that are different:
\begin{equation*}
n_{\alpha} = \textnormal{cardinality of } \{\alpha_1,\ldots, \alpha_K\}.
\end{equation*}
Notice that  $n_\alpha$  is also the number of noise samples $\{Z_{1,t}, \ldots, Z_{K,t}\}$ that are not exactly the same, but that differ by a constant factor not equal to 1. It is also equal to the number of different rows (or columns) in $\mat{K}_z$.

\begin{theorem}\label{th:K}
  If  all the noises are of positive variance \eqref{eq:amos100},
  and if the covariance matrix $\mat{K}_{z}$ has rank 1, then the
  prelog is $n_{\alpha}$:
\begin{equation}
\varlimsup_{P\to \infty} \frac{ C_{\textnormal{K-BC},\Sigma}(P,\mat{K}_z)}{\frac{1}{2} \log(1+P)} = n_\alpha.
\end{equation}
\end{theorem}
\begin{proof}
  Since there are only $n_\alpha$ different channels, the prelog
  cannot exceed $n_\alpha$. It thus only remains to prove achievability.

  If $n_\alpha=K$, then a prelog $n_\alpha$ is achievable using the
  scheme presented in Section~\ref{sec:SchemeK} ahead; see
  Proposition~\ref{lem:K} at the end of that
  section. 
  If $n_\alpha<K$, then we pick $n_\alpha$ receivers such that the
  corresponding $\alpha$'s are all different. We then apply the scheme
  in Section~\ref{sec:SchemeK} to only these $n_\alpha$ receivers (and
  ignore the other receivers). 
\end{proof}
\begin{corollary}
When $\mat{K}_z$ is of rank 1 and all its rows are different (i.e., $n_\alpha=K$), the prelog is $K$.
\end{corollary}

The achievability of prelog $K$, for $K\geq 3$, was proved in
\cite{ArdestMF:11} for the \emph{complex} memoryless Gaussian
broadcast channel.  In \cite{ArdestMF:11}, however,  it is assumed that
the real and imaginary parts of the noise symbols are
correlated. Consequently, Theorem~\ref{th:K} is not implied by
\cite{ArdestMF:11}.

\begin{note}
  Theorem~\ref{th:K} remains valid also when the transmitter has
  feedback only from a single receiver.  The proof is analogous to the
  proof of Note~\ref{th:note}.
\end{note}

\subsection{A Scheme for the case where $\textnormal{rank}(\mat{K}_z)=1$ and $n_\alpha=K$}\label{sec:SchemeK}
We generalize the coding scheme of Section~\ref{sec:proof-awgn-bc} to
the case where there are $K\geq 2$ receivers. We 
focus on the  case where $\mat{K}_z$ has rank~$1$ and
 \begin{equation}\label{eq:nalpha}
 n_\alpha =K,
 \end{equation}
 i.e., 
\begin{equation}\label{eq:alphadiff}
\alpha_k \neq \alpha_{k'}, \quad \Bigl(k, k' \in\{1,\ldots, K\}, \: k\neq k'\Bigr).
\end{equation}
 

The idea is to exploit the feedback in order to transform each block of $\eta$
uses of the original scalar BC into a single use of a new MISO BC with
vector input $\trans{(\Xi_1,\ldots, \Xi_K)}\in\Reals^K$ and scalar outputs
$I_1,\ldots, I_K$ at Receivers~$1,\ldots, K$, and to then code over
these blocks (ignoring the feedback). This allows us to achieve on the
original scalar BC the scaled-by-$\eta^{-1}$ capacity of the new MISO
BC (without feedback).

We next describe how to transform the first block of~$\eta$ uses of
the original BC into a single use of the new MISO BC; subsequent
$\eta$-length blocks are transformed similarly.  The key
parameters are: the $\eta$-by-$\eta$ strictly lower-triangular
matrices $\mat{B}_1, \ldots, \mat{B}_K$; the $\eta$-dimensional
column-vectors $\vect{u}_1, \ldots, \vect{u}_K$; and the
$\eta$-dimensional column-vectors $\vect{v}_1, \ldots,
\vect{v}_K$. How to choose these parameters will be discussed later.

The transmitter produces the $\eta$-length vector of inputs
\begin{equation}
\bfX=\sum_{k=1}^K \Xi_{k} \vect{u}_k +  \sum_{k=1}^K \mat{B}_k \alpha_k\bftZ_{1}
\end{equation}
where $\bfZ_{1}\triangleq\trans{(Z_{1,1}, \ldots, Z_{1,\eta})}$ are
the first $\eta$ samples of the noise experienced by Receiver~1, which
can be computed strictly-causally by the transmitter thanks to the
feedback.  Using~\eqref{eq:BCK1} and~\eqref{eq:alk}, we can express the
channel outputs $\bfY_{k}\triangleq \trans{(Y_{k, 1}, \ldots,
  Y_{k,\eta})}$ observed by Receiver~$k\in\{1,\ldots, K\}$ as
\begin{IEEEeqnarray}{rCl}
\bfY_{k}& =&\sum_{k'=1}^K \Xi_{k} \vect{u}_k +  \left( \sum_{k'=1}^K \mat{B}_{k'} \alpha_{k'} +\mat{I}\alpha_k\right) \bftZ_{1}.
\end{IEEEeqnarray}
Based on these outputs, Receiver~$k$ computes its new scalar output 
\begin{equation}\label{eq:IKk}
I_{k} \triangleq \trans{\vect{v}}_k \bfY_{k}.
\end{equation}
The channel input sequence satisfies the average block-power
constraint over the block of length $\eta$
whenever the information carrying symbols $\Xi_1, \ldots, \Xi_K$ are
independent; they satisfy $\E{\Xi_k^2}\leq1$ for every $k \in \{1,
\ldots, K\}$; and 
\begin{equation}\label{eq:powerblockK}
\sum_{k=1}^{K} \|\vect{u}_k\|^2 +  \trace{ \left( \sum_{k=1}^{K}  \mat{B}_k \alpha_k\right) { \left( \sum_{k=1}^{K}  \trans{\mat{B}_k}\alpha_k \right)}} \leq \eta P.
\end{equation}


For every integer $\eta\geq K$, we next present a choice of the
parameters $\mat{B}_1, \ldots, \mat{B}_K, \vect{u}_1, \ldots,
\vect{u}_K$, and $\vect{v}_1, \ldots, \vect{v}_K$ with the following
properties:
\begin{itemize}
\item[(i)] each vector  $\vect{v}_k$ is  orthogonal to the first $\eta-1$ columns of the matrix $\big(\sum_{k'=1}^{K}\mat{B}_{k'} \alpha_{k'} +\mat{I}\alpha_k\big)$;
\item[(ii)] each vector $\vect{u}_{k}$ is orthogonal to  the vectors $\vect{v}_{1}, \ldots, \vect{v}_{k-1}, \vect{v}_{k+1}, \ldots, \vect{v}_K$ but not to $\vect{v}_k$; 
\item[(iii)] each inner product $\trans{\vect{v}}_{k} \vect{u}_k$ is proportional to $\big(\sqrt{P}/\alpha_k\big)^{\eta-K}$, where the proportionality factor is nonzero and does not depend on $\eta$ or $P$; and 
\item[(iv)]  the power constraint \eqref{eq:powerblockK} is satisfied for all $P\geq K$.
\end{itemize}

Properties~(i) and (ii) guarantee that the new scalar output~$I_k$
formed at Receiver~$k$ has the form
\begin{IEEEeqnarray}{rCl}\label{eq:Ik}
I_{k} &= & \trans{\vect{v}}_k \vect{u}_k \Xi_{k} + \alpha_{k} {Z}_{1,\eta},
\end{IEEEeqnarray}
i.e., the first $\eta-1$ noise symbols $Z_{1,1},\ldots,$ $
Z_{1,\eta-1}$ and the interference symbols $\{\Xi_{k'}\}_{k'\neq k}$
are completely canceled
out. 
As explained in more detail later, Property~(iii) guarantees that, when $\eta\to \infty$, our scheme achieves prelog 1 to each Receiver~$k\in\set{K}$.

To describe our choice of the parameters, we need
definitions~\eqref{eq:alphak} and \eqref{eq:wk} ahead. 
Define for each $k\in\{1,\ldots, K\}$ the $K$-dimensional column-vector 
\begin{equation}\label{eq:alphak}
\boldsymbol{\alpha}_k = \trans{\begin{pmatrix} 1 & \alpha_k & \alpha_k^2 & \alpha_k^3 & \ldots & \alpha_k^K\end{pmatrix}}.
\end{equation}
Let $\vect{\hat{w}}_{k}$ be the projection of $\boldsymbol{\alpha}_k$
onto the linear subspace spanned by $\{\boldsymbol{\alpha}_1, \ldots,
\boldsymbol{\alpha}_{k-1}, \boldsymbol{\alpha}_{k+1},\ldots,
\boldsymbol{\alpha}_K\}$, and define 
\begin{equation}\label{eq:wk}
\vect{w}_k\triangleq\frac{ \boldsymbol{\alpha}_k- \vect{\hat{w}}_{k}}{ \left\|  \boldsymbol{\alpha}_k- \vect{\hat{w}}_{k}\right\|}.
\end{equation} 
Note that the vectors $\{\boldsymbol{\alpha}_{k}\}$ and
$\{\vect{w}_k\}$ do not depend on $P$ or $\eta$.

For every $\eta\geq K$, choose the matrices $\mat{B}_1,
\ldots, \mat{B}_K$ so that
\begin{IEEEeqnarray}{rCl}\label{eq:B}
\sum_{k'=1}^{K}  \mat{B}_{k'} \alpha_{k'}= 
 \begin{pmatrix}
0 & 0& \cdots & 0 & 0 &0\\
\sqrt{P} &   0& 0& \cdots & 0 &0&  \\
0& \sqrt{P}&0&   \cdots & 0 &0 \\

\vdots& \ddots& \ddots& \ddots& \vdots& \vdots
\\
 0 &  \cdots & 0 &\sqrt{P} & 0 &0\\
 0 & \cdots & 0&0&\sqrt{P}& 0
\end{pmatrix},\IEEEeqnarraynumspace
\end{IEEEeqnarray}
and choose  for $k\in\{1,\ldots,K\}$:
\begin{IEEEeqnarray}{rCl}
\vect{u}_k&=&\trans{\begin{pmatrix} \frac{w_{k,1}}{\sqrt{P}^{K-1}}  & \!\frac{w_{k,2}}{ \sqrt{P}^{K-2}} & \!\ldots & \!\frac{w_{k,K-1}}{\sqrt{P}}&\!w_{k,K} & \!0 &\!\ldots &\!0)\end{pmatrix}} \nonumber \\ \label{eq:uk}
\end{IEEEeqnarray} 
where $w_{k,j}$ denotes the $j$-th entry of the vector $\vect{w}_{k}$, and 
\begin{IEEEeqnarray}{rCl}
\vect{v}_k &=& \trans{ \begin{pmatrix}\left( -\frac{\sqrt{P}}{\alpha_k}\right)^{\eta-1} & \left( -\frac{\sqrt{P}}{\alpha_k}\right)^{\eta-2}& \ldots &-\frac{\sqrt{P}}{\alpha_k}&1 \end{pmatrix}}.\IEEEeqnarraynumspace\label{eq:vk}
\end{IEEEeqnarray}

We next verify that this parameter choice satisfies Properties
(i)--(iv). Property~(i) follows from~\eqref{eq:B} and
\eqref{eq:vk}. By~\eqref{eq:alphazero} and \eqref{eq:alphadiff} the
vectors $\{\boldsymbol{\alpha}_k\}_{k=1}^K$ are linearly
independent. Therefore, by~\eqref{eq:wk} and by the definition of the
vectors $\{\vect{\hat{w}}_k\}$, each vector
$\vect{w}_k$ 
is orthogonal to $\{\boldsymbol{\alpha}_1, \ldots,
\boldsymbol{\alpha}_{k-1}, \boldsymbol{\alpha}_{k+1}, \ldots,
\boldsymbol{\alpha}_{K}\}$ but not to
$\boldsymbol{\alpha}_k$: \begin{subequations}\label{eq:alphakwk}
 \begin{equation}
 \trans{\boldsymbol{\alpha}}_{k'} \vect{w}_{k} = 0, \qquad k'\in\{1,\ldots, k-1, k+1, \ldots, K\},
 \end{equation} 
and 
 \begin{equation}\label{eq:alphaknzero}
 \trans{\boldsymbol{\alpha}}_{k} \vect{w}_{k} \neq 0.
 \end{equation} 
 \end{subequations}
Property~(ii) follows now by~\eqref{eq:alphaknzero} and because by \eqref{eq:uk} and \eqref{eq:vk}:
\begin{equation}\label{eq:innervu}
\trans{\vect{v}}_{k'} \vect{u}_{k} = \frac{ \sqrt{P}^{\eta-K}}{\alpha_{k}^{\eta-1}} \trans{\boldsymbol{\alpha}}_{k'} \vect{w}_{k}, \qquad k, k'\in\{1,\ldots, K\}
\end{equation}
Property~(iii) follows by~\eqref{eq:alphakwk} and ~\eqref{eq:innervu},
and because $\boldsymbol{\alpha}_k$ and $\boldsymbol{w}_k$ do not
depend on $\eta$ or $P$.  Finally, the last Property~(iv) (the power
constraint for $P> K$) follows by combining~\eqref{eq:B} with
\eqref{eq:uk} and because, by definition~\eqref{eq:wk},
$\|\vect{w}_k\|^2=1$.

We conclude that the new output $I_k$ at Receiver~$k$ is  of the form in~\eqref{eq:Ik}. Therefore, by~\eqref{eq:innervu}, when  $P> K$ our  scheme with the described  choice of parameters achieves all rate tuples $(R_1,\ldots, R_K)$ that  satisfy
\begin{IEEEeqnarray}{rCl}\label{eq:Rk}
R_k& \leq  & \frac{1}{2\eta} \log \left( 1+ \frac{ P^{\eta-K}(\trans{\boldsymbol{\alpha}}_k\vect{w}_k)^2}{\alpha_k^{2\eta}}\right), \quad k\in\{1,\ldots, K\}, \nonumber \\ 
\end{IEEEeqnarray}
for some $\eta\geq K$. Since $\trans{\boldsymbol{\alpha}}_k\vect{w}_k$
is not zero~\eqref{eq:alphaknzero} and does not depend on $\eta$, 
by letting $\eta$ tend to infinity, 
we obtain from~\eqref{eq:Rk}:
\begin{proposition}\label{lem:K}
If rank$(\mat{K}_z)=1$ and $n_\alpha=K$ and if the power constraint $P>K$, then with noise-free feedback all nonnegative rate-tuples $(R_1,\ldots, R_K)$ satisfying 
\begin{IEEEeqnarray}{rCl}
R_k&< & \frac{1}{2} \log^+ \left(\frac{P}{\alpha_k^2}  \right), \quad k\in\{1,\ldots, K\},
\end{IEEEeqnarray}
are achievable over the $K$-user Gaussian BC.
\end{proposition}
Thus, for each Receiver~$k$, we can achieve prelog 1; and therefore we
achieve prelog $K$ for the sum-rate.

\section{Two-User Interference Channel}
\label{sec:inter}
\subsection{Setup and Results}

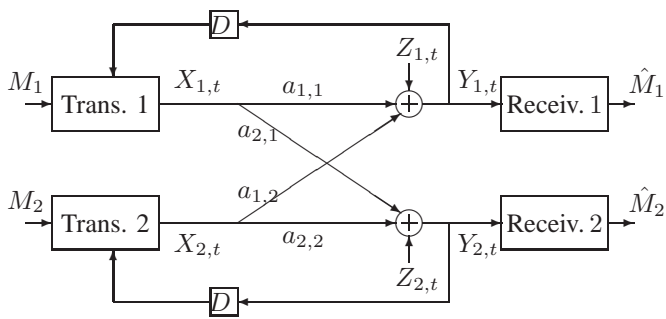
\begin{figure}[h!]
  \begin{center}
  \setlength{\unitlength}{1pt}
  \begin{picture}(230,100)(-30,50)
    \put (-30, 115) {\vector (1, 0) {10} }
    \put (-35, 118) {\makebox (10,10) {$M_1$}}
    \put (-30, 70) {\vector (1, 0) {10} }
    \put (-35, 73) {\makebox (10,10) {$M_2$}}

    \put (-20, 105) {\framebox (40,20) {{Trans. 1}}}

    \put (-20, 60) {\framebox (40,20) {{Trans. 2}}}

    \put (20, 115) {\vector (1, 0) {90} }
    \put (50, 115) {\vector (3, -2) {62} }
    \put (30, 118) {\makebox (10,10) {$X_{1,t}$}}
   \put (120, 115) {\vector (1, 0) {30} }
    \put (136, 118) {\makebox (10,10) {$Y_{1,t}$}}
    \put (115, 115) {\circle {10} }
    \put (112, 115) {\line (1, 0) {6} }
    \put (115, 112) {\line (0, 1) {6} }
    \put (115, 130) {\vector (0, -1) {10} }
    \put (113, 130) {\makebox (10,10) {$Z_{1,t}$}}

    \put (130, 115) {\line (0, 1) {30} }
    \put (130, 145) {\vector (-1, 0) {80} }
    \put (40, 140) {\framebox (10,10) {{$D$ }}}
    \put (40, 145) {\line (-1, 0) {37} }
    \put (03, 145) {\vector (0, -1) {20} }

    \put (20, 70) {\vector (1, 0) {90} }
    \put (50, 70) {\vector (3, 2) {62} }
    \put (30, 56) {\makebox (10,10) {$X_{2,t}$}}
    \put (120, 70) {\vector (1, 0) {30} }
    \put (136, 56) {\makebox (10,10) {$Y_{2,t}$}}
    \put (115, 70) {\circle {10} }
    \put (112, 70) {\line (1, 0) {6} }
    \put (115, 67) {\line (0, 1) {6} }
    \put (115, 55) {\vector (0, 1) {10} }
    \put (113, 43) {\makebox (10,10) {$Z_{2,t}$}}

      \put (70, 115) {\makebox (10,10) {$a_{1,1}$}}
       \put (53, 98) {\makebox (10,10) {$a_{2,1}$}}
        \put (53, 75) {\makebox (10,10) {$a_{1,2}$}}
         \put (70, 58) {\makebox (10,10) {$a_{2,2}$}}

    \put (130, 70) {\line (0, -1) {30} }
    \put (130, 40) {\vector (-1, 0) {80} }
    \put (40, 35) {\framebox (10,10) {{$D$ }}}
    \put (40, 40) {\line (-1, 0) {37} }
    \put (03, 40) {\vector (0, 1) {20} }

    \put (150, 105) {\framebox (40,20) {{Receiv. }$\!1$}}
    \put (190, 115) {\vector (1, 0) {10} }
    \put (200, 118) {\makebox (10,10) {$\hat{M}_1$}}

    \put (150, 60) {\framebox (40,20) {{Receiv. }$\!2$}}
    \put (190, 70) {\vector (1, 0) {10} }
    \put (200, 73) {\makebox (10,10) {$\hat{M}_2$}}

  \end{picture}
  \end{center}
  \vspace{4mm}
  
  \caption{The two-user  Gaussian IC with one-sided noise-free feedback.}
  \label{Fig-ICnfgen}
\end{figure}

In this section we study the real scalar memoryless Gaussian IC with noise-free
feedback, which is depicted in Figure~\ref{Fig-ICnfgen}. This network
has two transmitters and two receivers: Transmitter~1 wishes to send
Message~$M_1$ to Receiver~1, and Transmitter~2 wishes to send
Message~$M_2$ to Receiver~2.  Assuming that at time~$t$ Transmitter~1
sends the real symbol $x_{1,t}$ and Transmitter~2 sends the real
symbol $x_{2,t}$, Receiver~1 observes
\begin{subequations}\label{eq:channelY}
\begin{eqnarray}
Y_{1,t} & = & a_{1,1}x_{1,t} + a_{1,2} x_{2,t} +Z_{1,t},
\end{eqnarray} 
and  Receiver~2 observes
\begin{eqnarray} 
Y_{2,t} & = & a_{2,1}x_{1,t}+a_{2,2}x_{2,t} +Z_{2,t}.
\end{eqnarray}
\end{subequations}
The channel gains are non-zero real constants
\begin{equation}
a_{1,1}, a_{1,2}, a_{2,1}, a_{2,2}\neq 0,
\end{equation}
and the noise sequences $\{(Z_{1,t},Z_{2,t})\}_{t=1}^n$ are as
in Section~\ref{sec:pfb}.

Each transmitter has access to noise-free feedback from its
intended receiver. Thus, each transmitter can choose its time-$t$ channel input as 
\begin{eqnarray*}
X_{k,t}&=&f^{(n)}_{\textnormal{IC},k,t}\left( M_k, {Y}_k^{t-1}
\right), \quad k\in\{1,2\},
\end{eqnarray*}
for some encoding function $f_{\IC,k,t}^{(n)}$ of the form 
\begin{IEEEeqnarray*}{rCl}
f_{\IC,k,t}^{(n)} \colon \mathcal{M}_{k} \times \mathbb{R}^{t-1}
\to \Reals,
\quad \nu \in\{1,2\}.
\end{IEEEeqnarray*} 
The two channel input sequences are subject to the same average block-power
constraint  $P>0$:
\begin{eqnarray}
\frac{1}{n} \E{  \sum_{t=1}^n X_{k,t}^2  } \leq P, \quad k\in\{1,2\}.\label{eq:powerIC}
\end{eqnarray}
Decoding rules, achievable rate pairs, the capacity region, the
sum-rate capacity, and the prelog are defined as for the Gaussian BC. We
denote the sum-rate capacity of the Gaussian IC with noise-free feedback
by $C_{\textnormal{IC}, \Sigma}(P, \sigma_1^2, \sigma_2^2, \rho_z)$.


Without feedback, the prelog of the Gaussian IC equals 1; with noise-free
feedback it can be 2, depending on the channel gains $a_{1,1}, a_{1,2}, a_{2,1}, a_{2,2} \neq 0$ and on the noise parameters $\sigma_1^2, \sigma_2^2, \rho_z$.
\begin{theorem}\label{th:Thic}
  The prelog of the Gaussian IC with noise-free feedback satisfies 
  the following three statements.
\begin{itemize}
\item If $|\rho_z| < 1$ or if $|\rho_z| = 1$ and
  $\frac{a_{2,2}}{a_{1,2}}= \frac{a_{1,2}}{a_{1,1}}=
  \rho_z\frac{\sigma_2}{\sigma_1}$, then
\begin{equation}\label{eq:rhoopenic} \varlimsup_{P \rightarrow
      \infty} \frac{C_{\textnormal{IC},\Sigma}(P, \sigma_1^2, \sigma_2^2, \rho_z)}{\frac{1}{2} \log ( 1 + P )} = 1;
\end{equation}
\item if $|\rho_z|=1$ and neither $\frac{a_{2,2}}{a_{1,2}}$ nor
  $\frac{a_{1,2}}{a_{1,1}}$ equals $\rho_z
  \frac{\sigma_2}{\sigma_1}$, then
  \begin{equation}
    \label{eq:rho-1ic} 
    \varlimsup_{P \rightarrow
      \infty} \frac{C_{\textnormal{IC},\Sigma}(P, \sigma_1^2, \sigma_2^2, \rho_z)}{\frac{1}{2} \log ( 1 + P )} = 2;
\end{equation}
\item otherwise 
\begin{equation}\label{eq:rel1ic}
1\leq   \varlimsup_{P \rightarrow \infty} \frac{C_{\textnormal{IC},\Sigma}(P, \sigma_1^2, \sigma_2^2, \rho_z)}{\frac{1}{2} \log ( 1 + P )} \leq 2.
\end{equation} 
\end{itemize}
\end{theorem}
\begin{proof}
See Section~\ref{sec:PICFB}.
\end{proof}


\subsection{A Scheme}\label{sec:scheme_ic}
We present a scheme similar to the scheme for the BC in
Section~\ref{sec:Scheme}.  Thus, the idea is to transform each block
of $\eta$ channel uses of the original IC into a single use of a new
IC with inputs $\Xi_1\in \Reals$ at Transmitter~1 and $\Xi_2 \in
\Reals$ at Transmitter~2 and with outputs $I_1$ at Receiver~1 and $I_2$ at
Receiver~2, and to then code over this new IC. In this way we can
achieve on the original IC the scaling-by-$\eta^{-1}$ of any rate pair
that is achievable on the new IC.

We describe how to transform the first block of $\eta$ uses of the
original IC into a single use of the new IC; subsequent blocks are
transformed similarly. The key parameters are:
   two  strictly lower-triangular  $\eta$-by-$\eta$ matrices $\mat{B}_1$ and $\mat{B}_2$; two $\eta$-dimensional column-vectors $\vect{u}_1, \vect{u}_2$; and two $\eta$-dimensional row-vectors $\vect{v}_1, \vect{v}_2$.

Denoting by
 \begin{equation} 
 \bfX_{k} \triangleq \trans{ ( X_{k, 1}, \ldots, X_{k,\eta} )}, \qquad k\in\{1,2\},
 \end{equation} 
 the $\eta$-length vector of symbols that Transmitter~$k$ sends in this first block, 
 we choose
 \begin{subequations}
\begin{IEEEeqnarray}{rCl}\label{ICinputs}
\bfX_{1}& =&\vect{u}_1  \Xi_{1} +   \mat{B}_1(a_{1,2}\bfX_{2}+ \bftZ_{1})\\
\bfX_{2}& =& \vect{u}_2  \Xi_{2} + \mat{B}_2(a_{2,1}\bfX_{1} + \bftZ_{2}).
\end{IEEEeqnarray}
\end{subequations}
Receiver~$k$ observes the corresponding $\eta$-length vector of outputs $\vect{Y}_k$ 
and computes the new output
\begin{IEEEeqnarray}{rCl}
I_{k}& =&\trans{\vect{v}}_k   \bfY_{k}.
\end{IEEEeqnarray}

In the following, we present a choice of parameters for the case where $\rho_z\in\{-1,1\}$. In this case, 
\begin{equation}
Z_{2,t}= \rho_z \frac{\sigma_2}{\sigma_1} {Z}_{1,t},
\end{equation} and we
 can rewrite the channel outputs at Receiver~2 as 
 \begin{subequations}\label{eq:Yeq}
 \begin{equation}
Y_{2,t} = a_{2,1}X_{1,t} + a_{2,2} X_{2,t}  +\rho_z \frac{\sigma_2}{\sigma_1} {Z}_{1,t},
\end{equation}
\end{subequations}

We choose $\mat{B}_2$ the all-zero matrix and 
\begin{IEEEeqnarray}{rCl}\label{eq:B1}
\mat{B}_1
& =& \begin{pmatrix}
0 & 0& 0& \cdots  & 0 &0\\
\frac{\sqrt{P}}{\sigma_1} &   0&0 & \cdots & 0& 0 \\
0& \frac{\sqrt{P}}{\sigma_1} &  0&\cdots & 0 & 0 & \\
\vdots& \ddots& \ddots& \ddots&\vdots & \vdots
\\
 0 &\ldots &  0&\frac{\sqrt{P}}{\sigma_1}  & 0 &0\\
 0 &0 & \ldots &0 &\frac{\sqrt{P}}{\sigma_1} & 0
\end{pmatrix}.\IEEEeqnarraynumspace
\end{IEEEeqnarray}
By \eqref{ICinputs} and \eqref{eq:Yeq}, with this choice
\begin{IEEEeqnarray*}{rCl}
\bfY_{1}&=& a_{1,1}\vect{u}_1 \Xi_{1} +  a_{1,2}(a_{1,1}\mat{B}_1 +\mat{I}) \vect{u}_2\Xi_{2}\nonumber \\
&& +  (a_{1,1}\mat{B}_1+\mat{I}) \bftZ_{1}
\end{IEEEeqnarray*}
and
\begin{IEEEeqnarray*}{rCl}
\bfY_{2}& = & a_{2,1} \vect{u}_1 \Xi_{1} + (a_{2,1}a_{1,2}\mat{B}_1 +a_{2,2}\mat{I}) \vect{u}_2\Xi_{2} \nonumber \\
&&+ \left(a_{2,1} \mat{B}_1+\rho_z \frac{\sigma_2}{\sigma_1}\mat{I}\right) \bftZ_{1}. 
\IEEEeqnarraynumspace
\end{IEEEeqnarray*}
We now choose the vector $\vect{v}_1$ to be orthogonal to the first $(\eta-1)$ columns of the matrix $(a_{1,1}\mat{B}_1+\mat{I})$;  $\vect{v}_{2}$ to be orthogonal to the first $(\eta-1)$ columns of the matrix $ \big(a_{2,1} \mat{B}_1+\rho_z \frac{\sigma_2}{\sigma_1}\mat{I}\big) $; the vector $\vect{u}_1$ to be orthogonal to $\vect{v}_2$ but not to $\vect{v}_1$; and the vector $\vect{u}_2$ simply not orthogonal to $ (a_{2,1}a_{1,2}\mat{B}_1 +a_{2,2}\mat{I})\vect{v}_2$. Such a choice is:
\begin{IEEEeqnarray*}{rCl}
\vect{u}_1 & = & \sqrt{\frac{P/2}{1+a_{2,1}^2P/\sigma_2^2 }} \cdot \trans{\begin{pmatrix} 1 & \frac{a_{2,1}\sqrt{P}  }{\rho_z\sigma_2} & 0 \ldots & 0\end{pmatrix}}\\
\vect{u}_2 & = & \sqrt{\frac{\sigma_1^2}{2a_{1,2}^2}} \cdot \trans{\begin{pmatrix} 1& 0 & \ldots &0 \end{pmatrix}}\\
\vect{v}_1 & = & \begin{pmatrix} \left(\frac{-a_{1,1}\sqrt{P}}{\sigma_1} \right)^{\eta-1} &  \left(\frac{-a_{1,1}\sqrt{P}}{\sigma_1} \right)^{\eta-2} & \ldots &  \frac{-a_{1,1}\sqrt{P}}{\sigma_1} &1 \end{pmatrix}\\
\vect{v}_2 & = & \begin{pmatrix} \left(\frac{-a_{2,1}\sqrt{P}  }{\rho_z\sigma_2}\right)^{\eta-1} &  \left(\frac{-a_{2,1}\sqrt{P} }{\rho_z\sigma_2}\right)^{\eta-2}  & \ldots &  \frac{-a_{2,1}\sqrt{P} }{\rho_z\sigma_2}&1 \end{pmatrix}.
\end{IEEEeqnarray*}
If the information symbols $\Xi_1$ and $\Xi_2$ are independent and
satisfy $\E{\Xi_1^2},\E{\Xi_2^2}\leq1$, then the chosen $\vect{u}_1,
\vect{u}_2, \mat{B}_1, \mat{B}_2$ result in a scheme satisfying the
blocklength-$\eta$ average power constraint at Transmitter 1 for any
positive integer $\eta$ and also at Transmitter~2 for $\eta>
\frac{\sigma_1^2}{2P}$. 

We obtain:
\begin{subequations}\label{eq:IIC}
\begin{IEEEeqnarray}{rCl}\lefteqn{
I_{1} \triangleq \trans{\vect{v}}_1 \vect{Y}_{1}}\nonumber \\&=& \sqrt{\frac{P\sigma_2^2}{\sigma_2^2+Pa_{2,1}^2 }} \Bigg(\!-\frac{\sqrt{P}a_{1,1}}{\sigma_1}\Bigg)^{\!\eta-1}\!\!\bigg( 1-  \frac{a_{2,1}\sigma_1}{a_{1,1}\rho_z\sigma_2}\bigg)a_{1,1}\Xi_{1} \nonumber \\
&&+ {Z}_{1,\eta}
\label{eq:IIC1}
\end{IEEEeqnarray}
and 
\begin{IEEEeqnarray}{rCl}\label{eq:IIC2}
\lefteqn{
I_{2}\triangleq \trans{\vect{v}}_2 \vect{Y}_{2}} \nonumber \\ &=&\Bigg(-\frac{\sqrt{P} a_{2,1} }{\rho_z\sigma_2}\Bigg)^{\eta-1}\!\!\bigg( a_{2,2} - a_{1,2}\frac{\rho_z \sigma_2}{\sigma_1}\bigg) \Xi_{2}+ \rho_z\frac{\sigma_2}{\sigma_1} {Z}_{1,\eta}.\nonumber\\
\end{IEEEeqnarray}
\end{subequations}
Thus,  the noise symbols $Z_{1,1}, \ldots, Z_{1, \eta-1}$ 
are completely canceled out when forming the "new outputs" in~\eqref{eq:IIC1} and \eqref{eq:IIC2} and only $Z_{1, \eta}$ 
remains.
Moreover, the "interference symbol" $\Xi_{2}$ is canceled out in $I_{1}$ and the "interference symbol" $\Xi_{1}$ is canceled out in $I_{2}$. 

By~\eqref{eq:IIC}, we conclude that our scheme achieves all nonnegative rate pairs that satisfy 
\begin{subequations}\label{eq:RatesetaIC}
\begin{IEEEeqnarray}{rCl}
R_1 &< & \frac{1}{2\eta} \log\left( 1+ \frac{P^{\eta}a_{1,1}^{2\eta}}{\sigma_1^{2\eta}}\frac{\sigma_2^2}{(\sigma_2^2+Pa_{2,1}^2) } \left(1- \frac{a_{2,1}\sigma_1}{a_{1,1}\rho_z\sigma_2}\right)^2\right)\nonumber \\\\
R_2 & <  & \frac{1}{2\eta} \log \left(1+ \frac{P^{\eta-1} a_{2,1}^{2\eta-2}}{\sigma_2^{2\eta}} \left( a_{2,2} - \frac{a_{1,2}\rho_z \sigma_2}{\sigma_1}\right)^2\right)
\end{IEEEeqnarray}
\end{subequations}
for all $\eta\geq \frac{\sigma_1^2}{2P}$. 

Taking the limit $\eta\to\infty$ leads to the following.
\begin{proposition}\label{lem:ic}
  If $\rho_z\in\{-1,1\}$ and neither $\frac{a_{2,1}}{\alpha_{1,1}}$
  nor $\frac{a_{2,2}}{a_{1,2}}$ equals
  $\rho_z\frac{\sigma_2}{\sigma_1}$, then all rate-pairs satisfying
\begin{IEEEeqnarray}{rCl}
R_1 & < & \frac{1}{2} \log^+ \left( \frac{a_{1,1}^2P}{\sigma_1^2} \right)\\
R_2 & < & \frac{1}{2} \log^+ \left( \frac{a_{2,1}^2  P}{\sigma_2^2} \right)
\end{IEEEeqnarray}
are achievable over the Gaussian IC with noise-free feedback.
\end{proposition}

\begin{note}
  With the proposed choice of parameters our scheme achieves prelog 2
  when $\rho_z \in \{-1,1\}$ and $\frac{a_{2,1}}{\alpha_{1,1}}$ and
  $\frac{a_{2,2}}{a_{1,2}}$ are both different from
  $\rho_z\frac{\sigma_2}{\sigma_1}$ and when $\eta \to \infty$.  For
  fixed $\eta$ the scheme achieves a prelog of $2
  \frac{\eta-1}{\eta}$; see the rate constraints
  in~\eqref{eq:RatesetaIC}. Thus, choosing $\eta=3$ suffices to
  achieve a prelog larger than 1.
\end{note}

\begin{note}\label{eq:remic}
Exchanging the roles of the two transmitters we obtain: when $\rho_z\in\{-1,1\}$ and  $\frac{a_{2,1}}{\alpha_{1,1}}$ and $\frac{a_{2,2}}{a_{1,2}}$ are both different from $\frac{\rho_z\sigma_2}{\sigma_1}$, then all rate-pairs satisfying
\begin{IEEEeqnarray}{rCl}
R_1 & \leq & \frac{1}{2} \log^+ \left( \frac{a_{1,2}^2P}{\sigma_1^2} \right)\\
R_2 & \leq & \frac{1}{2} \log^+ \left( \frac{a_{2,2}^2  P}{\sigma_1^2} \right)
\end{IEEEeqnarray}
are achievable over the Gaussian IC with noise-free feedback.
\end{note}

\begin{note}
For a symmetric setup where $a_{1,1}=a_{2,2}$ and $a_{1,2}=a_{2,1}$  the achievability of \eqref{eq:rho-1ic}  can also be shown using a slight generalization of Kramer's memoryless LMMSE scheme \cite{Kramer:02}, see \cite{GastparLW:10}.
\end{note}

\newcommand{\Q}[1]{\mathcal{Q} \left( #1\right)}

 \subsection{Proof of Theorem~\ref{th:Thic}}\label{sec:PICFB} 

Relation~\eqref{eq:rel1ic} follows from the following more general result: Irrespective of the channel parameters,
\begin{equation}\label{eq:trivial2}
1 \leq   \varlimsup_{P \rightarrow \infty} \frac{C_{\textnormal{IC},\Sigma}(P, \sigma_1^2, \sigma_2^2, \rho_z)}{\frac{1}{2} \log ( 1 + P )} \leq 2.
\end{equation}
The lower bound in \eqref{eq:trivial2} can be achieved by silencing
Transmitter~1 and letting Transmitter~2 communicate its Message $M_2$
to Receiver~2 over the resulting interference-free Gaussian channel
$Y_{2,t}=a_{2,2}X_{2,t}+Z_{2,t}$ at rate
$R_2=\frac{1}{2}\log\left(1+\frac{a_{2,2}^2P}{\sigma_2^2}
\right)$. The upper bound can be derived using the cut-set bound and
the entropy maximizing property of the Gaussian distribution under a
covariance matrix constraint. In fact, applying two cuts between both
transmitters and each of the two receivers yields the following upper
bounds
\begin{eqnarray*}
R_k < \frac{1}{2} \log\left(1+\frac{(|a_{k,1}|+|a_{k,2}|)^2P}{\sigma_k^2} \right), \quad k\in\{1,2\},
\end{eqnarray*}
which establish the converse result in \eqref{eq:trivial2}.

We next prove~\eqref{eq:rhoopenic}. The achievability follows
from~\eqref{eq:trivial2}. When $\rho_z\in\{-1,1\}$ and
$\frac{a_{2,1}}{a_{1,1}}=\frac{a_{2,2}}{a_{1,2}}=\rho_z\frac{\sigma_2}{\sigma_1}$,
the converse holds because in this case
\begin{IEEEeqnarray}{rCl}
Y_{1,t}=\rho_z\frac{\sigma_1}{\sigma_2} Y_{2,t} \quad \textnormal{with probability }1
\end{IEEEeqnarray}
and thus, each receiver can reconstruct the other receiver's
outputs. Consequently, the feedback capacity of our Gaussian IC coincides
with the feedback capacity of the Gaussian MAC from the two transmitters
to one of the two receivers, and its prelog is 1 \cite{Ozarow:84}.

To prove the converse to \eqref{eq:rhoopenic} when $\rho_z\in(-1,1)$ we use a genie-argument and a
generalized Sato-MAC bound \cite{Sato:81}, similar to the upper bounds
in \cite[Section V-B]{HanK:81}, \cite{Kramer:04, LapidothShamaiWigger:07itw, LapidothLSW:09}. Our proof
consists of the following three steps.  In the first step
we let a genie  reveal the symbols 
\[U^n=Z_2^n-\frac{a_{2,2}}{a_{1,2}}Z_1^n
\] to Receiver~1 before the transmission begins. This obviously can
only increase the sum-rate capacity of our channel. We refer to the
resulting setup as the \emph{genie-aided IC}.

In the second step, we apply Sato's MAC-bound argument \cite{Sato:81}
to this {genie-aided IC}.\footnote{Unlike in Sato's setup, here both transmitters have
  feedback from their corresponding receivers. However, as we shall see, also in our setup (because the feedback is one-sided) we can use the same arguments.} That is, we define an appropriate \emph{genie-aided MAC} and  show that the capacity of the genie-aided IC is contained in the capacity  of this genie-aided MAC. 
The genie-aided MAC is obtained from the genie-aided IC by eliminating Receiver~2 and requiring that the sole remaining Receiver~1
decode both messages $M_1$ and $M_2$. The desired inclusion of the capacities is proved by showing that for any encoding and decoding
strategies for the genie-aided IC it is possible to find encoding/decoding strategies for the genie-aided MAC such that the probability of error over the MAC is no larger than over the IC. 

Given encoding/decoding functions for the genie-aided IC, we choose
the encoding/decoding functions for the genie-aided MAC as
follows. The MAC transmitters apply the same encoding functions as the
IC transmitters. The sole MAC-receiver decodes the pair $(M_1,M_2)$ as
follows: 1.) It applies IC-Receiver~1's decoding rule to decode
Message $M_1$. 2.) It computes
\begin{eqnarray}
\hat{X}_{1,t} = f_{\IC,1,t}^{(n)}(\hat{M}_1, Y_1^{t-1}),\qquad t\in\{1,\ldots,n\},
\end{eqnarray}
and 
\begin{eqnarray}\label{eq:funcoutputs}
\hat{Y}_{2}^n = \frac{a_{2,2}}{a_{1,2}} \big({Y}_{1}^n- a_{1,1}\hat{X}_{1}^n\big)+a_{2,1}\hat{X}_1^n+ U^n,
\end{eqnarray}
where  $\hat{M}_1$ denotes the decoded message in 1.) and $f_{\IC,1,t}^{(n)}$ denotes IC-Transmitter~1's encoding function.
3.) It finally  applies IC-Receiver~2's decoding rule to decode Message $M_2$ based on the sequence $\hat{Y}_2^n$. 

Notice that if the MAC-receiver (and thus also IC-Receiver 1) decodes
$M_1$ correctly, then $\hat{X}_1^n=X_1^n$, and $\hat{Y}_2^n=Y_2^n$,
and the MAC's guess of $M_{2}$ is identical to that of the
IC's. Consequently, whenever the IC-Receivers~1 and 2 decode their
intended messages $M_1$ and $M_2$ correctly, so does the sole
MAC-receiver, and the probability of error over the MAC cannot
therefore exceed the probability of error over the IC. This concludes
the second step.
 
In the third step we show that the genie-aided MAC has prelog no
larger than 1.  Combined with the previous two steps this yields the
desired converse to \eqref{eq:rhoopenic}.  Before elaborating
on this third step, we recall that in the genie-aided MAC the channel
law is 
\[
Y_{1,t}=a_{1,1}X_{1,t}+a_{1,2} X_{2,t}+Z_{1,t}, \quad t\in\{1,\ldots, n\};
\]
the two transmitters observe the generalized feedback signals
$\{Y_{1,t}\}$ and $\{Y_{2,t}\}$; and before the transmission begins,
the receiver learns the genie-information $U^n$. 

We now prove that the prelog of this genie-aided MAC is upper-bounded by 1. To
this end we fix an arbitrary sequence of blocklength-$n$, rates-$(R_1,R_2)$
coding schemes for the considered MAC such that the probability of
error $\epsilon(n)$ tends to zero as $n$ tends to infinity. For
every blocklength $n$ we then have:
\allowdisplaybreaks[4]
\begin{eqnarray}\lefteqn{
R_1+R_2} \nonumber\\ & \leq &\frac{1}{n} I(M_1,M_2; {Y}_1^n , {U}^n) +\frac{\epsilon(n)}{n}\nonumber\\
 & = & \frac{1}{n} I(M_1,M_2; {Y}_{1}^n|{U}^n) +\frac{\epsilon(n)}{n} \nonumber\\
& = & \frac{1}{n} \sum_{t=1}^n \Big(h(Y_{1,t}|{Y}_1^{t-1},
{U}^n) \nonumber\\ & & \quad \quad \quad  - h(Y_{1,t}|{Y}_1^{t-1}, M_1,M_2,
{U}^n) \Big) +\frac{\epsilon(n)}{n}\nonumber\\
& \leq  & \frac{1}{n} \sum_{t=1}^n \Big(h(Y_{1,t}|U_{t})\nonumber\\ & & \quad
 \quad \quad - h(Y_{1,t}|
 {Y}_1^{t-1}, M_1,M_2, {Y}_{2}^{t-1},
{U}^n) \Big) +\frac{\epsilon(n)}{n} \nonumber\\ 
& = & \frac{1}{n} \sum_{t=1}^n \Big( h(Y_{1,t}|U_{t})- h(Y_{1,t}|
 X_{1,t},X_{2,t}, U_{t}) \Big)\nonumber\\
& = & \frac{1}{n} \sum_{t=1}^n I(Y_{1,t};X_{1,t},X_{2,t}|U_{t})\nonumber \\
& \leq & \frac{1}{2} \log \left( 1+ \frac{(|a_{1,1}|+|a_{1,2}|)^2P}{ \Var{ Z_{1,t}| Z_{2,t} - \frac{a_{2,2}}{a_{1,2}} Z_{1,t}}}\right)\label{eq:lastIC}
\end{eqnarray} 
where the first inequality follows by Fano's inequality; the first
equality follows by the independence of the genie-information ${U}^n$
and the messages $M_1$ and $M_2$; the third equality by noting that
the vector ${Y}_2^{t-1}$ can be computed as a function of $M_1,
{Y}_1^{t-1}$, and ${U}^{t-1}$, see \eqref{eq:funcoutputs}; the fourth
equality follows because the input $X_{1,t}$ is a function of the
Message $M_1$ and the feedback outputs $Y_1^{t-1}$, and similarly
$X_{2,t}$ is a function of $M_2$ and $Y_2^{t-1}$,  and
because of the Markov relation \[(M_1,M_2, {Y}_1^{t-1}, {Y}_{2}^{t-1},
{U}^{t-1}, {U}_{t+1}^n) - (X_{1,t},X_{2,t},U_t) - Y_{1,t};\] and the
last inequality follows because the Gaussian distribution maximizes
differential entropy under a covariance constraint. 

Since $\Var{Z_{1,t}|Z_{2,t}-\frac{a_{2,2}}{a_{1,2}} Z_{1,t}}$ does not depend on $P$ and is strictly positive whenever $\rho_z\in(-1,1)$, 
by \eqref{eq:lastIC} 
we conclude that \eqref{eq:rhoopenic} holds also when $\rho_z\in(-1,1)$.

The converse to \eqref{eq:rho-1ic} follows from the general
Relation~\eqref{eq:trivial2}, and its achievability from  Proposition~\ref{lem:ic}  in Section~\ref{sec:scheme_ic}.

\section*{Acknowledgment}

The authors thank Prof. Frans M. J. Willems, TU Eindhoven, for
pointing them to~\cite{WillemsM:81}, which inspired the investigation
leading to this work. They also thank the Associate Editor and the anonymous reviewers for their valuable comments.

\appendices

\section{Proof of Theorem~\ref{thm:asympt}}\label{sec:pt2}
Recall that here $\rho_z(P)$ depends on the power $P$. 

 The following two upper bounds are obtained from the cut-set bound and the fact that  a Gaussian law maximizes the differential entropy under a variance constraint \cite{CoverT:06}. With two individual cuts between the transmitter and each of the two receivers we obtain
\begin{IEEEeqnarray}{rCl}
R_1+ R_ 2 & \leq&  \max_{X \colon \E{X^2} \leq P} \{I(X;Y_1) +I(X;Y_2)\}\nonumber\\
& = & \frac{1}{2} \log\left( 1+ \frac{P}{\sigma_1^2}\right)+ \frac{1}{2} \log\left( 1+ \frac{P}{\sigma_2^2}\right), \label{eq:cut1}
\end{IEEEeqnarray}and with a single cut between the transmitter and both receivers
\begin{IEEEeqnarray}{rCl}
R_1+ R_ 2 & \leq&  \max_{X \colon \E{X^2} \leq P} I(X;Y_1,Y_2) \nonumber\\
& \leq & \frac{1}{2} \log\left( 1+ \frac{P(\sigma_1^2+\sigma_2^2- 2\sigma_1\sigma_2 \rho_z )}{ \sigma_1^2\sigma_2^2(1-\rho_z^2) }\right). \IEEEeqnarraynumspace \label{eq:cut2}
\end{IEEEeqnarray}

We first prove the converse to \eqref{eq:same} where $\sigma_1^2=\sigma_2^2$.
In this case, Upper bound \eqref{eq:cut2}  specializes to 
\begin{IEEEeqnarray}{rCl}\label{eq:uprhoopen}
\lefteqn{C_{\BCFB,\Sigma}(P, \sigma_1^2, \sigma_1^2, \rho_z(P))} \quad \nonumber \\ & \leq &  \frac{1}{2}\log \left( 1+ \frac{P}{ \frac{\sigma_1^2}{2} (1+\rho_z(P))}\right), \quad \rho_z(P)\in(-1,1).\IEEEeqnarraynumspace
\end{IEEEeqnarray}
In view of \eqref{eq:BCphysdeg}, and since we assume $\sigma_1^2=\sigma_2^2$ and  we defined $-\log(0)=\infty$, Upper bound \eqref{eq:uprhoopen} holds also for $\rho_z(P)\in\{-1,1\}$, and thus  for all  $\rho_z(P)\in[-1,1]$. Therefore, by the definition of $\zminus$ in \eqref{eq:conda}, 
 \begin{IEEEeqnarray}{rCl} \label{eq:u142}
\varlimsup_{P\rightarrow \infty} \frac{
C_{\BCFB,\Sigma}(P, \sigma_1^2, \sigma_1^2, \rho_z(P))}{\frac{1}{2} \log(1+ P)}
&\leq &   1+ \zminus.
\end{IEEEeqnarray}
On the other hand,  by \eqref{eq:cut1}, irrespective of  $\{\rho_z(P)\}_{\{P>0\}}$,
\begin{IEEEeqnarray}{rCl}\label{eq:u141}
\varlimsup_{P\rightarrow \infty} \frac{
C_{\BCFB,\Sigma}(P, \sigma_1^2, \sigma_1^2, \rho_z(P))}{\frac{1}{2} \log(1+ P)} \leq 2.
\end{IEEEeqnarray}
Combining \eqref{eq:u142} and \eqref{eq:u141} establishes the converse to \eqref{eq:same}.

We now prove the converse to \eqref{eq:alb} where $\sigma_1^2\neq \sigma_2^2$. Using the facts that 
\begin{equation}
\sigma_1^2 +\sigma_2^2 -2\rho_z(P)\sigma_1 \sigma_2< 2(\sigma_1^2 +\sigma_2^2)
\end{equation} and  
\begin{equation}1-\rho_z^2(P)\geq 1-|\rho_z(P)|,
\end{equation}
 we can further upper bound the right-hand side of \eqref{eq:cut2} to obtain:
\begin{IEEEeqnarray}{rCl}
\lefteqn{
C_{\BCFB,\Sigma}(P, \sigma_1^2, \sigma_2^2, \rho_z(P))}\; \nonumber \\
 &\leq & \frac{1}{2} \log \left( 1+
 \frac{P}{\frac{\sigma_1^2\sigma_2^2}{2(\sigma_1^2 +\sigma_2^2)}
  (1-|\rho_z(P)|)}\right), \;\; \rho_z(P)\in(-1,1).
     \nonumber\\\label{eq:C45}
\end{IEEEeqnarray} 
Now, since we defined $-\log(0)=\infty$, Upper bound \eqref{eq:C45} holds also for $\rho_z(P)\in\{-1,1\}$, and hence for all $\rho_z(P)\in[-1,1]$. Moreover, since by the   definitions of $\zminus$ and $\zplus$ in \eqref{eq:conda} and \eqref{eq:condb},
 \begin{equation}\label{eq:limmax}
 \varlimsup_{P\to \infty} \frac{ -\log( 1- |\rho_z(P)|)}{\log(P)} = \max \left\{\zminus,\zplus\right\},
 \end{equation}
 we  conclude that 
 \begin{IEEEeqnarray}{rCl} \label{eq:u143}
\varlimsup_{P\rightarrow \infty} \frac{
C_{\BCFB,\Sigma}(P, \sigma_1^2, \sigma_2^2, \rho_z(P))}{\frac{1}{2} \log(1+ P)}
&\leq &   1+ \max\left\{\zminus,\zplus\right\}.\IEEEeqnarraynumspace
\end{IEEEeqnarray} 
Combining  \eqref{eq:u143}  with \eqref{eq:u141} establishes the  converse to \eqref{eq:alb}.

We next prove that 
for arbitrary $\sigma_1^2, \sigma_2^2$:
 \begin{IEEEeqnarray}{rCl} \label{eq:l141}
\varlimsup_{P\rightarrow \infty} \frac{
C_{\BCFB,\Sigma}(P, \sigma_1^2, \sigma_2^2, \rho_z(P))}{\frac{1}{2} \log(1+ P)}
&\geq &  \min\left\{ 1+  \zminus,2\right\}. \IEEEeqnarraynumspace
\end{IEEEeqnarray} 
Since a generalized prelog of 1 is achievable even without  feedback \cite{Bergmans:74,Cover:72}
the interesting case is  $\zminus>0$.
In the following, assume that $\zminus>0$, which implies 
the existence of an increasing unbounded sequence $\{P_{\ell}\}_{\ell=1}^\infty$ such that   
  \begin{IEEEeqnarray}{rCl}\label{eq:limzminus}
  \lim_{\ell\to\infty} \frac{-\log(1+ \rho_z(P_\ell))}{\frac{1}{2} \log(P_\ell)} = \zminus>0,
  \end{IEEEeqnarray}
  and  in particular 
  \begin{equation}\label{eq:limitzminus}
  \lim_{\ell \to \infty} \rho_z(P_\ell)=-1.
  \end{equation}

For each $\ell$ we choose parameters $q_\ell$ and $\delta_\ell$ and show that Inequality \eqref{eq:l141} follows from Corollary~\ref{th:rates} specialized to these parameters. 
We choose 
\begin{IEEEeqnarray}{rCl}\label{eq:deltal}
\delta_\ell  =\begin{cases}  \frac{\sigma_1}{\sigma_2} \cdot \frac{\sigma_1-\rho_z(P_\ell) \sigma_2}{\sigma_2 - \rho_z(P_\ell) \sigma_1} & \textnormal{ if } \rho_z(P_\ell)\in(-1,1)\\
-\rho_z(P_\ell)\frac{\sigma_1}{\sigma_2}   & \textnormal{ if } \rho_z(P_\ell)\in\{-1,1\},
\end{cases}
\end{IEEEeqnarray}
and define the limit (not necessarily finite)
\begin{IEEEeqnarray}{rCl}
\kappa&\triangleq&\varlimsup_{\ell\to \infty} P_\ell \big(\sigma_1^2 +\delta^2_\ell\sigma_2^2+2\delta_\ell \rho_z(P_\ell) \sigma_1\sigma_2\big). 
\end{IEEEeqnarray} 
Depending on $\kappa$, we choose $q_\ell>0$ as follows. Let $\epsilon\in(0,1)$ be a small positive number.  
\begin{itemize}
\item 
If $\kappa=\infty$,  we choose 
\begin{equation}\label{eq:kappainfty}
q_\ell= \left( \frac{(1-\epsilon)P_\ell}{ \delta_\ell^2(1+\delta_\ell)^2 \left(\sigma_1^2 +\sigma_2^2\delta_\ell^2 +2\rho_z(P_\ell) \delta_\ell \sigma_1 \sigma_2\right)} \right)^{1/4};
\end{equation}
\item if $\kappa \in [0, \infty)$, we 
choose 
\begin{equation}\label{eq:kappa}
q_\ell= \big( \beta (1-\epsilon)P_\ell\big)^{1/2}
\end{equation}
where $\beta>0$ is a solution to
\begin{equation}\label{eq:condbeta}
\sigma_1^2 \left(1+\frac{\sigma_1}{\sigma_2}\right)^2\beta\left( 1+\beta \frac{\kappa}{\sigma_2^2}\right)=1.
\end{equation} 
\end{itemize}
Notice that for every $\epsilon>0$ there exists a positive integer $\ell_0(\epsilon, \sigma_1^2, \sigma_2^2, \kappa)$ such that our choice $(\delta_\ell, q_\ell)$ satisfies the power constraint~\eqref{eq:bpower} for all $\ell>\ell_0(\epsilon, \sigma_1^2, \sigma_2^2, \kappa)$. 

Moreover, if $\kappa\in[0,\infty)$, then specializing the rates in Corollary~\ref{th:rates} to the choices in  \eqref{eq:deltal} and \eqref{eq:kappa} proves that  \begin{IEEEeqnarray}{rCl}\label{eq:beforelast}
\varlimsup_{\ell\rightarrow \infty} \frac{
C_{\BCFB,\Sigma}(P_\ell, \sigma_1^2, \sigma_2^2, \rho_z(P_\ell))}{\frac{1}{2} \log(1+ P_\ell)}
&\geq &  2.\IEEEeqnarraynumspace
\end{IEEEeqnarray} 
If $\kappa=\infty$, then for all sufficiently large $\ell$ the correlation coefficient $\rho_z(P_\ell) \in(-1,1)$, and by \eqref{eq:deltal} the choice in \eqref{eq:kappainfty} evaluates to 
\begin{IEEEeqnarray}{rCl}\label{eq:newql}
q_\ell & =& 
\left( \frac{(\sigma_2-\sigma_1\rho_z(P_\ell))^2}{ \delta_\ell^2(1+\delta_\ell)^2 \sigma_1^2\left(\sigma_2^2+\sigma_1^2 - 2 \rho_z(P_\ell) \sigma_1 \sigma_2\right) }\right)^{1/4}\nonumber \\
& & \cdot \left(\frac{ (1-\epsilon)P_\ell}{ 1-|\rho_z(P_\ell)|^2} \right)^{1/4}
\end{IEEEeqnarray}
Notice that by  \eqref{eq:limitzminus} and \eqref{eq:deltal}
\begin{equation} \label{eq:alphanoteq}
\lim_{\ell\to \infty} \frac{\log\left(\frac{(\sigma_2-\sigma_1\rho_z(P_\ell))^2}{\delta_\ell^2(1+\delta_\ell)^2 \sigma_1^2\left(\sigma_2^2+\sigma_1^2 - 2 \rho_z(P_\ell) \sigma_1 \sigma_2\right)(1+|\rho_z(P_\ell)|)}\right)}{\frac{1}{2}\log(P)}=0, 
\end{equation}
 and therefore, specializing the rates in Proposition~\ref{th:rates} to the choice in \eqref{eq:deltal} and \eqref{eq:newql} proves that
 \begin{IEEEeqnarray}{rCl}\label{eq:last}
\varlimsup_{\ell\rightarrow \infty} \frac{
C_{\BCFB,\Sigma}(P_\ell, \sigma_1^2, \sigma_2^2, \rho_z(P_\ell))}{\frac{1}{2} \log(1+ P_\ell)}
&\geq &  1+\zminus.\IEEEeqnarraynumspace
\end{IEEEeqnarray} 
Combining \eqref{eq:last} with \eqref{eq:beforelast}  establishes \eqref{eq:l141}.

In a similar way we can also prove that when $\sigma_1^2\neq \sigma_2^2$ 
 \begin{IEEEeqnarray}{rCl} \label{eq:l142}
\varlimsup_{P\rightarrow \infty} \frac{
C_{\BCFB,\Sigma}(P, \sigma_1^2, \sigma_2^2, \rho_z(P))}{\frac{1}{2} \log(1+ P)}
&\geq &  \min\left\{ 1+  \zplus,2\right\}. \IEEEeqnarraynumspace
\end{IEEEeqnarray} 
To this end, it suffices that in the  proof to \eqref{eq:l141} we replace $\zminus$ by $\zplus$;  \eqref{eq:limzminus} by 
 \begin{IEEEeqnarray}{rCl}
  \lim_{\ell\to\infty} \frac{-\log(1- \rho_z(P_\ell))}{\frac{1}{2} \log(P)} = \zplus>0;
  \end{IEEEeqnarray} 
   \eqref{eq:limitzminus} by 
   \begin{equation}\lim_{\ell\to\infty} \rho_z(P)=1;
   \end{equation} and \eqref{eq:condbeta} by 
   \begin{IEEEeqnarray}{rCl}\label{eq:condbeta2}
\sigma_1^2 \left(1-\frac{\sigma_1}{\sigma_2}\right)^2\beta\left( 1+\beta \frac{\kappa}{\sigma_2^2}\right)=1.
   \end{IEEEeqnarray} 
  The assumption of non-equal noise variances $\sigma_1^2\neq \sigma_2^2$ is needed here to conclude that \eqref{eq:alphanoteq} holds and that \eqref{eq:condbeta2} has a finite solution for $\beta$.

Combining finally \eqref{eq:l142} with \eqref{eq:l141} establishes the achievability of \eqref{eq:same} and \eqref{eq:alb} and concludes the proof.

\section{Proof of Note~\ref{rem:noisyfb}}\label{sec:Noisylim}
Let $\sigma_1^2, \sigma_2^2>0$ and $\rho_z\in[-1,1]$ be fixed and for every power $P>0$ let the feedback-noise variances $\sigma_{W1}^2(P)$ and $\sigma_{W2}^2(P)$ be given, where
\begin{equation}\label{fbnoiselim}
\varlimsup_{P\to \infty} \frac{ - \log (\sigma_{W\nu}^2(P))}{\log P} \leq 0, \qquad \nu\in\{1,2\}.
\end{equation}

Since for each power $P$ a prelog 1 is achievable even without feedback
we have to prove 
\begin{equation}\label{eq:rr}
\varlimsup_{P\to \infty} \frac{ C_{\textnormal{BCNoisy},\Sigma}\big(P,\sigma_1^2, \sigma_2^2, \rho_z,
\sigma_{W1}^2(P), \sigma_{W2}^2(P)\big)}{ \frac{1}{2}\log( 1+P)} \leq 1.
\end{equation}
For $\rho_z \in (-1,1)$  Inequality \eqref{eq:rr} follows immediately from Corollary~\ref{th:Th},  because with noisy feedback the prelog cannot be larger than with noise-free feedback. (The transmitter can always add the feedback noise itself.)

For $\rho_z\in\{-1,1\}$ the proof of \eqref{eq:rr} is similar to the proof in Section~\ref{sec:PNoisyFB}. In fact, following the same steps as before, we can conclude that for each $P>0$
\begin{IEEEeqnarray}{rCl}\lefteqn{
C_{\textnormal{BCNoisy},\Sigma}(P,\sigma_1^2, \sigma_2^2, \rho_z,
\sigma_{W1}^2(P), \sigma_{W2}^2(P)) } \nonumber\qquad \\ & \leq &\frac{1}{2} \log\left( 1+ \frac{ P
  }{\min\left\{ \Var{{Z}_{1,t}'}, \Var{{Z}_{2,t}'} \right\}
  }\right),\label{eq:noisyup}\IEEEeqnarraynumspace
\end{IEEEeqnarray}
where here, because $|\rho_z|=1$, the variances in \eqref{eq:Var1} and \eqref{eq:Var2} simplify to
\begin{IEEEeqnarray}{rCl}
\Var{{Z}_{1,t}'} & = &  \frac{\sigma_1^2 \sigma_{W1}^2(P)\sigma_{W2}^2(P) }{\sigma_{1}^2\sigma_{W2}^2(P)+ \sigma_2^2\sigma_{W1}^2(P)+\sigma_{W1}^2(P)\sigma_{W2}^2(P) }, \nonumber \\ \IEEEeqnarraynumspace \label{eq:Var1t}\\
\Var{{Z}_{2,t}'} & = &   \frac{\sigma_2^2 \sigma_{W1}^2(P)\sigma_{W2}^2(P) }{\sigma_{1}^2\sigma_{W2}^2(P)+ \sigma_2^2\sigma_{W1}^2(P)+\sigma_{W1}^2(P)\sigma_{W2}^2(P) }.\nonumber \\\IEEEeqnarraynumspace \label{eq:Var2t}
\end{IEEEeqnarray}
The desired inequality \eqref{eq:rr}  for $\rho_z \in\{-1,1\}$ follows now simply by combining \eqref{fbnoiselim} with \eqref{eq:noisyup}--\eqref{eq:Var2t}.

\section{Proof of Lemma~\ref{lem:opteta}}\label{app:lemma}
Let $\eta \in \mathbb{Z}^+$. If $1+\zeta\geq \xi$, then
\begin{IEEEeqnarray*}{rCl}
\frac{1}{2\eta} \log( 1+  \xi^{\eta-1} \zeta)& \leq &\frac{1}{2\eta} \log( 1+  (1+\zeta)^{\eta-1} \zeta) \\ &= &\frac{1}{2\eta} \log\Big( (1+\zeta)^{\eta} -\underbrace{\left( (1+\zeta)^{\eta-1}-1\right)}_{\geq 0}\Big)\\
& \leq & \frac{1}{2\eta} \log\big( (1+\zeta)^{\eta} \big)\\
& = & \frac{1}{2} \log(1+ \zeta).
\end{IEEEeqnarray*}
Otherwise, if $1+\zeta < \xi$, then
\begin{IEEEeqnarray*}{rCl}
\frac{1}{2\eta} \log( 1+  \xi^{\eta-1} \zeta)& \leq & \frac{1}{2\eta} \log( 1+  \xi^{\eta-1} (\xi-1))\\ 
&= &\frac{1}{2\eta} \log\Big( \xi^{\eta} -\underbrace{\left( \xi^{\eta-1}-1\right)}_{\geq 0}\Big)\\
& \leq & \frac{1}{2\eta} \log\big( \xi^{\eta} \big)\\
& = & \frac{1}{2} \log(\xi).
\end{IEEEeqnarray*}

\bibliographystyle{ieeetr}

\begin{IEEEbiographynophoto}{Michael Gastpar}
received the Dipl. El.-Ing. degree from ETH Z\"urich, in 1997, the M.S. degree from the University of Illinois at Urbana-Champaign, Urbana, IL, in 1999, and the
Doctorat \`es Science degree from Ecole Polytechnique F\'ed\'erale (EPFL), Lausanne, Switzerland, in 2002, all in electrical engineering. He was also a student in engineering and philosophy at the Universities of Edinburgh and Lausanne.

He is a Professor in the School of Computer and Communication
Sciences, Ecole Polytechnique F\'ed\'erale (EPFL), Lausanne, Switzerland.
He was an Assistant (2003-2008) and tenured Associate Professor (2008-2011)
with the Department of Electrical Engineering and Computer Sciences,
University of California, Berkeley, where he still holds a faculty position.
He also holds a faculty position at Delft University of Technology, The Netherlands,
and was a Researcher with the Mathematics of Communications Department,
Bell Labs, Lucent Technologies, Murray Hill, NJ.
His research interests are
in network information theory and related coding and signal processing techniques,
with applications to sensor networks and neuroscience.

Dr. Gastpar won the 2002 EPFL Best Thesis Award, an NSF CAREER Award
in 2004, an Okawa Foundation Research Grant in 2008, and an ERC Starting Grant in 2010.
He is the co-recipient of the 2013 Communications Society \& Information Theory Society Joint Paper Award. He was an Information Theory Society Distinguished Lecturer (2009Ð2011), an Associate Editor for Shannon Theory for the {\sc IEEE Transactions on
   Information Theory} (2008-2011), and he has served as Technical Program Committee Co-Chair for the
2010 International Symposium on Information Theory, Austin, TX.
\end{IEEEbiographynophoto}

\begin{IEEEbiographynophoto}{Amos Lapidoth}
 (S'89, M'95, SM'00, F'04) received the B.A.\ degree in
 mathematics ({\em summa cum laude}, 1986), the B.Sc.\  degree in
 electrical engineering ({\em summa cum laude}, 1986), and the M.Sc.\
 degree in electrical engineering (1990) all from the
 Technion---Israel Institute of Technology. He received the Ph.D.\
 degree in electrical engineering from Stanford University in 1995.

 In the years 1995--1999 he was an Assistant and Associate Professor
 at the Department of Electrical Engineering and Computer Science at
 the Massachusetts Institute of Technology, and was the KDD Career
 Development Associate Professor in Communications and Technology.
 He is now Professor of Information Theory at ETH Zurich in
 Switzerland. He is the author of the book \emph{A Foundation in
 Digital Communication}, published by Cambridge University Press in
 2009. His research interests are in digital communications and
 information theory.

 Dr.\ Lapidoth served in the years 2003--2004 and 2009 as Associate
   Editor for Shannon Theory for the {\sc IEEE Transactions on
   Information Theory.}
\end{IEEEbiographynophoto}

\begin{IEEEbiographynophoto}{Yossef Steinberg}
(MÕ96ÐSMÕ09, F'11) received the B.Sc., M.Sc., and Ph.D. degrees
in electrical engineering in 1983, 1986, and 1990, respectively, all from Tel-Aviv
University, Tel-Aviv, Israel.
He was a Lady Davis Fellow in the Department of Electrical Engineering,
Technion---Israel Institute of Technology, Haifa, Israel, and held visiting appointments
in the Department of Electrical Engineering at Princeton University,
Princeton, NJ, and at the C~I~Center, George Mason University, Fairfax, VA.
From 1995 to 1999, he was with the Department of Electrical Engineering,
Ben Gurion University, Beer-Sheva, Israel. In 1999, he joined the Department
of Electrical Engineering at the Technion.
Dr. Steinberg served in the years 2004--2007 as Associate Editor for Shannon Theory, 
and currently serves as Associate Editor at large, for the 
{\sc IEEE Transactions on Information Theory.} 
Dr. Steinberg's research interests are in Digital Communications, Information Theory,
and Estimation.
He won the 2007 Information Theory Society Paper Award,
jointly with Hanan Weingarten and Shlomo Shamai.
\end{IEEEbiographynophoto}

\begin{IEEEbiographynophoto}{Mich\`ele Wigger}
(S'05-M'09) received the M.Sc. degree in electrical
engineering (with distinction) and the Ph.D. degree in electrical engineering
both from ETH Zurich in 2003 and 2008, respectively. In 2009 she was
a postdoctoral researcher at the ITA center at the University of California,
San Diego. Since December 2009 she is an Assistant Professor at Telecom
ParisTech, in Paris, France. 
Her research interests are in information and
communications theory.
\end{IEEEbiographynophoto}

\end{document}